\documentclass[fleqn,usenatbib]{mnras}

\usepackage{newtxtext,newtxmath}

\usepackage[T1]{fontenc}

\DeclareRobustCommand{\VAN}[3]{#2}
\let\VANthebibliography\thebibliography
\def\thebibliography{\DeclareRobustCommand{\VAN}[3]{##3}\VANthebibliography}

\usepackage{graphicx}	
\usepackage{amsmath}	
\usepackage[dvipsnames]{xcolor}
\usepackage{orcidlink}
\usepackage{hyperref}
\usepackage{ulem}

\title[CHF with ML]{Exploring the Dependence of Gas Cooling and Heating Functions on the Incident Radiation Field with Machine Learning}

\author[D. Robinson et al.]{
David Robinson \orcidlink{0000-0002-3751-6145},$^{1}$\thanks{E-mail: dbrobins@umich.edu}
Camille Avestruz \orcidlink{0000-0001-8868-0810},$^{1,2}$
Nickolay Y. Gnedin \orcidlink{0000-0001-5925-4580}$^{3,4,5}$
\\
$^{1}$Department of Physics; University of Michigan, Ann Arbor, MI 48109, USA\\
$^{2}$Leinweber Center for Theoretical Physics; University of Michigan, Ann Arbor, MI 48109, USA\\
$^{3}$Theoretical Physics Division; 
Fermi National Accelerator Laboratory;
Batavia, IL 60510, USA\\
$^{4}$Kavli Institute for Cosmological Physics;
The University of Chicago;
Chicago, IL 60637, USA\\
$^{5}$Department of Astronomy \& Astrophysics; 
The University of Chicago; 
Chicago, IL 60637, USA
}

\date{Accepted XXX. Received YYY; in original form ZZZ}

\pubyear{2023}

\begin{document}
\label{firstpage}
\pagerange{\pageref{firstpage}--\pageref{lastpage}}
\maketitle

\begin{abstract}
Gas cooling and heating functions play a crucial role in galaxy formation. But, it is computationally expensive to exactly compute these functions in the presence of an incident radiation field. These computations can be greatly sped up by using interpolation tables of pre-computed values, at the expense of making significant and sometimes even unjustified approximations. Here, we explore the capacity of machine learning to approximate cooling and heating functions with a generalized radiation field. Specifically, we use the machine learning algorithm XGBoost to predict cooling and heating functions calculated with the photoionization code Cloudy at fixed metallicity, using different combinations of photoionization rates as features. We perform a constrained quadratic fit in metallicity to enable a fair comparison with traditional interpolation methods at arbitrary metallicity. We consider the relative importance of various photoionization rates through both a principal component analysis (PCA) and calculation of SHapley Additive exPlanation (SHAP) values for our XGBoost models. We use feature importance information to select different subsets of rates to use in model training.  Our XGBoost models outperform a traditional interpolation approach at each fixed metallicity, regardless of feature selection. At arbitrary metallicity, we are able to reduce the frequency of the largest cooling and heating function errors compared to an interpolation table. We find that the primary bottleneck to increasing accuracy lies in accurately capturing the metallicity dependence. This study demonstrates the potential of machine learning methods such as XGBoost to capture the non-linear behavior of cooling and heating functions.
\end{abstract}




\section{Introduction}
\label{intro}

Galaxy formation involves many interacting processes, of which the primary one is the gravitational collapse of baryonic gas into the potential wells of dark matter halos after the decoupling of baryons and photons.  Since baryonic gas can provide thermal pressure support, the rate at which the gas can dissipate energy (via cooling) is a critical factor in determining the density and temperature at which the gravitational collapse stops \citep{rees_ostriker77}. Cooling and heating functions are a traditional way of describing how the internal energy of the gas changes due to radiative processes \citep[e.g.][]{cox_tucker69, sutherland_dopita93, lykins13, wang14}, and determine the thermal evolution of gas \citep[e.g.][]{dalgarno72, gnat_sternberg07}. Coupled with other relevant processes, cooling and heating functions help determine the overall evolution of the gas \citep[e.g.][]{martinez_serrano08, richings14, galligan19, romero21}. Hence, cooling and heating functions are important to theoretical modeling of galaxy formation \citep[see the review of][]{benson10}. For example, the comparison between the gas cooling time and the gravitational freefall time introduces characteristic scales where the two are equal \citep[e.g.][]{rees_ostriker77, silk77, white_frenk91, kauffmann93}.  Cooling also regulates how efficiently additional gas can accrete onto a (proto)galaxy \citep[e.g.][]{binney77, bertschinger85, cole94, croton06, brooks09}, and the radiation emitted by the accreting gas \citep{fardal01}. Gas mass can be lost from a (forming) galaxy due to heating by ionizing radiation from outside the galaxy \citep{okamoto08}.

Given an atomic and molecular composition, and using values for relevant atomic properties, the equilibrium populations of various ionization states and energy levels in a gas can be computed  \citep[e.g.][]{spitzer62, arnaud85, ferland93, ferland09}. These populations can then be used to determine the cooling and heating rates of the gas \citep[e.g.][]{cox_tucker69, dalgarno72, sutherland_dopita93, ferland98}. In the absence of external sources of ionizing radiation, these populations are set by collisional ionization equilibrium (CIE) and become well-known and relatively simple functions of gas density, metallicity, and temperature only \citep{cox_tucker69, sutherland_dopita93}.

An incident radiation field can change the distribution of ionization states and energy level populations, and hence can alter cooling and heating functions from the CIE limit \citep[e.g.][]{wiersma09, gnedin_hollon12, galligan19, robinson22}. Computing cooling and heating functions for an arbitrary radiation field is a highly complex and computational expensive effort, requiring major development efforts and sophisticated software, in particular photoionization codes \citep[see the review of][]{kallman01} such as Cloudy \citep{ferland97, ferland98, ferland13, ferland17, chatzikos23}. Photoionization codes require a comprehensive database of atomic data related to ionization and recombination, such as XSTAR \citep{bautista01}. Photoionization codes have been developed for a variety of geometries, such as a spherical cloud with an illuminating source at the center \citep{kallman_mccray82}, a slab illuminated from either side \citep{dumont00}, a cone illuminated from the apex \citep{kinkhabwala03}, pseudo-3D geometries with non-extended sources \citep{morisset05}, and fully general 3D geometries \citep[e.g.][]{ercolano03, wood04, baes05}.

In this paper, we aim to accurately predict cooling and heating functions for an atomic gas in the presence of an incident radiation field, which may include a local component in addition to the extragalactic background.  Local contributions can dominate the radiation field inside galaxies and in the circumgalactic medium \citep{draine78}.

Since gas cooling (and heating) is a vital ingredient in galaxy formation, cooling and heating functions need to be included in numerical simulations of galaxy formation. However, the full calculation of level and ionization state populations is much too complex (and, hence, slow) for modeling cooling and heating functions on the fly in galaxy formation simulations. Hence, for simulation use, the cooling and heating functions must be approximated with a less computationally intensive method. One traditional approach to this is pre-computing cooling and heating functions with Cloudy on a grid of relevant parameters, and interpolating between table values in the simulation \citep[e.g.][]{kravtsov03, smith08, hopkins11, vogelsberger14}.  Cooling and heating tables for gas in CIE generally include dimensions of temperature, density, and metallicity \citep[e.g.][]{cox_tucker69, sutherland_dopita93, smith08}. The same table dimensions can be used for gas with a constant ionizing background at fixed redshift \citep[e.g.][]{hopkins11}. Incorporating a spatially constant UV background adds one additional dimension, the redshift $z$ (or radiation field strength), to the tables \citep[e.g.][]{kravtsov03, robertson_kravtsov08, smith17, gutcke21, schaye23}. In particular, various simulations incorporate the redshift-dependent cooling tables of \citet{wiersma09}, including simulations described in \citet{thomas09, schaye14, mccarthy17}. More recent simulations, such as those of \citet{gutcke21} and \citet{schaye23} incorporate the cooling tables of \citet{ploeckinger_schaye20}. These tables include an interstellar radiation field with a fixed frequency dependence and normalization depending on the gas column density, in addition to a spatially constant UV background. Tables with more dimensions can also be constructed for gas illuminated by an incident radiation field which includes local contributions from a synthesized stellar spectrum and a quasar-like power law \citep{gnedin_hollon12}. 

Interpolation tables are not the only approach to approximating cooling and heating functions. One alternate approach is simplified chemical networks which follow only the most relevant atoms and molecules, which \textit{can} be evaluated on-the-fly \citep[e.g.][]{anninos97, grassi11, richings14, salz15, bovino16}. Other works use Cloudy to calculate cooling and heating functions for gas properties sampled from a numerical galaxy formation simulation of interest \citep[e.g.][]{robertson_kravtsov08, wiersma10, romero21, robinson22}. These `true' cooling and heating functions can then be used to train a neural network to predict cooling and heating functions (at least for gas properties which are reasonably consistent with those seen in the training data) \citep[e.g.][]{grassi11, galligan19}.

Cooling and heating functions depend on several gas parameters, as well as the incident radiation field, which is in general a function of frequency. The effect of the incident radiation field on the gas can be described by the photoionization rates for all possible ionization states of elements in the gas. Since an interpolation table in very large dimension is computationally impractical, we need to find a smaller number of photoionization rates that are still able to capture the dependence of cooling and heating functions on the incident radiation field.  This is an example of the general class of dimensionality reduction problems.  Such problems are common in machine learning, and several machine learning techniques exist to approach these problems \citep[see reviews by][]{zebari20, jia22}.

Several additional factors which we do not include in this paper can also impact gas cooling and heating, such as cosmic rays, dust, and molecules.  Cosmic rays can ionize gas particles, heating the gas.  Dust grains contribute to the heating function via photoheating, and to cooling via collisions between gas particles and dust grains \citep{ferland93}.  Dust also impacts the cooling function indirectly, via the accretion of gas-phase metals onto dust grains.  In general, this accretion will modify the relative abundances of metals \citep{dwek98}. Note that UV radiation fields can destroy dust grains through sublimation \citep{guhathakurta_draine89}.  Hence, low incident ionizing radiation is needed to have significant dust mass. Molecules contribute to heating via photodissociation, and cooling via vibrational and rotational line emission \citep{richings14}. In a hydrodynamic simulation, additional numerical approximations for the contributions of molecules, dust, and cosmic rays to the total cooling and heating functions could be added to the atomic cooling and heating approximations described in this paper.

In this work, we use the machine learning algorithm eXtreme Gradient Boosting (XGBoost, \citet{chen_guestrin16}) to model the dependence of cooling and heating functions on gas properties and various radiation field parameters. We train XGBoost models on tabulated cooling and heating functions calculated with Cloudy \citep{ferland98}. XGBoost is known to perform well on this type of tabular training data \citep{shwartz-ziv_armon21, grinztajn22}. We also explore the SHapley Additive eXplanation (SHAP, \citet{lundberg_lee17}) values \citep{lundberg18, lundberg20} of our model inputs, which we use to determine what radiation field parameters are most predictive of cooling and heating functions. The XGBoost algorithm has been used in the astrophysics literature for a variety of tasks in classification \citep[e.g.][]{tamayo16, ivanov21, lucey22, luo22}, such as identifying pulsar candidates in gamma ray data \citep[e.g.][]{mirabal16, wang19} and separating stars, galaxies, and quasars in photometric data \citep[e.g.][]{jin19, chang21, fu21, golob21, li21, nakoneczny21, hughes22}; and regression \citep[as here, see also][]{calderon_berlind19, hayden20, machado21, dang22, andrae23}, including for predicting quasar redshifts from photometry \citep[so called "photo-z"s, e.g.][]{jin19, nakoneczny21, kunsagi-mate22}. Some works include the use of SHAP values to analyze feature importances \citep[e.g.][]{machado21, heyl23}.

In this paper, we first describe our methodology, including the Cloudy calculations we use to train and evaluate XGBoost models, the input features we use and their distributions and correlations, how we determine which combinatos of radiation field features to use, and how we compare and evaluate XGBoost models.  Next, we present results for trained models compared to each other and the interpolation table in \citet{gnedin_hollon12}, on both the training data grid and an independent off-grid sample.  Finally, we discuss our conclusions and potential future directions. 

\section{Data and Methods}
\label{method}

\subsection{Model Data Input}\label{method:data}

In order to train machine learning models to predict cooling and heating functions, we use computations from the photoionization code Cloudy \citep{ferland98}. The training data used here is very similar to that used in \citet{gnedin_hollon12}, with some additional information included.

For a given parcel of gas, we define cooling and heating functions through equation~(\ref{eq:chf_def}): 
\begin{equation}
    \left. \frac{dU}{dt} \right\vert_\mathrm{rad} = n_b^2 [\Gamma(T, \ldots) - \Lambda(T, \ldots)],
    \label{eq:chf_def}
\end{equation}
where $U$ is the thermal energy density of the gas, $n_b = n_\mathrm{H} + 4n_\mathrm{He} + \ldots$ is the number of density of baryons (protons and neutrons), $T$ is the gas temperature, and $\Gamma, \Lambda$ are, respectively, the heating and cooling functions. The prefactor $n_b^2$ is included because of the importance of collisional two-body processes in radiative cooling and heating.  In the limit of collisional ionization equilibrium (CIE), where there is no external ionizing radiation field on the gas, and accounting for two-body processes only, $\Lambda$ and $\Gamma$ are independent of $n_b$.  However, even in CIE, multi-electron processes such as Auger ionization and dielectronic recombination can contribute to the cooling and heating functions when the gas contains metals \citep{ferland98}, introducing $n_b$ dependence to $\Lambda$ and $\Gamma$.  Here, we consider the more general case of an external radiation field, and account for other relevant physical processes beyond two-body collisions.  Hence, $\Lambda$ and $\Gamma$ also depend on $n_b$, metallicity, the incident radiation field, and may depend on other physical characteristics - this complex dependence is captured by the dots in equation~(\ref{eq:chf_def}).

The input data we use consists of Cloudy computations of $\Lambda$ and $\Gamma$ evaluated on a grid of values for the gas properties from \citet{gnedin_hollon12}. The relevant gas properties used are temperature $T$, hydrogen number density $n_H$ \citep[since Cloudy uses the hydrogen number density $n_H$ as its density parameter instead of the baryon density $n_b$,][]{ferland98}, metallicity $Z$, and 4 parameters describing a model for the incident radiation field:

\begin{equation}
    J_\nu = J_0 \left[\frac{1}{1+f_Q} s_\nu + \frac{f_Q}{1+f_Q}x^{-\alpha} \right]e^{-\tau_\nu},
    \label{eq:rad_field}
\end{equation}
where $x = {h\nu}/({1 \mathrm{Ry}})$ is the photon energy in Rydbergs. The parameter $J_0$ in equation~(\ref{eq:rad_field}) accounts for the overall amplitude of the radiation field.  The expression in square brackets contains two terms: a stellar spectrum $s_\nu$ and an AGN-like power law with slope $\alpha$.  The parameter $f_Q$ quantifies the ratio of the stellar and AGN-like contributions to $J_\nu$ at 1 Ry.  The stellar spectrum $s_\nu$ is given by:
\begin{equation}
    s_\nu = \begin{cases}
    5.5 & x < 1, \\
    x^{-1.8} & 1 < x < 2.5, \\
    0.4x^{-1.8} & 2.5 < x < 4, \\
    2 \times 10^{-3} \frac{x^3}{e^{x/1.4} - 1} & x > 4, \\
    \end{cases}
    \label{eq:stellar_spec}
\end{equation}
and is a fit to stellar spectra from the Starburst99 spectral synthesis library \citep{starburst99}. The radiation field $J_\nu$ in equation~(\ref{eq:rad_field}) is also subject to attenuation due to neutral hydrogen and helium with optical depth $\tau_\nu$ given by:
\begin{equation}
    \tau_\nu = \frac{\tau_0}{\sigma_{\mathrm{HI}, 0}} \left[0.76\sigma_\mathrm{HI}(\nu) + 0.06\sigma_\mathrm{HeI}(\nu) \right],
    \label{eq:opt_depth}
\end{equation}
where $\tau_0$ describes the overall optical depth, $\sigma_\mathrm{HI, 0}$ is the photoionization cross section of neutral hydrogen at its ionization threshold, and $\sigma_\mathrm{HI}(\nu), \sigma_\mathrm{HeI}(
\nu)$ are the photoionization cross sections for neutral hydrogen and helium, respectively, as functions of frequency. A loose justification for this parameterization choice is that a random place in the universe may be irradiated by both stellar and AGN radiation, and the strongest nearby source may happen to lie behind a sufficiently dense absorbing cloud. More complex scenarios with multiple nearby sources of approximately equal strengths can obviously be imagined, but they are unlikely to be common.

In total, the input data we use is fully described by 7 parameters: $T, n_H, Z, J_0, f_Q,\tau_0$, and $\alpha$.  We define our training data on a grid of these parameters, described in Table~\ref{tab:training_data}.

\begin{table}
    \centering
    \begin{tabular}{ll}
         \hline
         Parameter & Values \\
         \hline 
         $\log{(T / \mathrm{K})}$ & $1, 1.1, 1.2, \ldots, 9,$ \\
         $\log{(n_H / \mathrm{cm}^{-3})}$ & $-6, -5, -4, \ldots, 6,$ \\
         $Z/Z_\odot$ & $0, 0.1, 0.3, 1, 3,$ \\
         $\log{(J_0 \, \mathrm{cm}^{-3} / n_b / J_\mathrm{MW})}$ & $-5, -4.5, -4, \ldots, 7,$ \\
         $\log{f_Q}$ & $-3, -2.5, -2, \ldots, 1,$ \\
         $\log{\tau_0}$ & $-1, -0.5, 0, \ldots, 3,$ \\
         $\alpha$ & $0, 0.5, 1, \ldots, 3,$ \\
         \hline
    \end{tabular}
    \caption{The parameters describing the training data table and the values they take.  Here, $J_\mathrm{MW} = 10^6 \, \mathrm{photons} \, \mathrm{cm}^{-2} \, \mathrm{s}^{-1} \, \mathrm{ster}^{-1} \, \mathrm{eV}^{-1}$ \citep{gnedin_hollon12}. The $Z=0$ case actually uses $Z = 10^{-4} Z_\odot$.}
    \label{tab:training_data}
\end{table}

The 4 radiation field parameters ($J_0, f_Q, \tau_0$, and $\alpha$) are only well-defined for a radiation field with the functional form given by equation~(\ref{eq:rad_field}), and cannot necessarily be determined for a given radiation field from a galaxy formation simulation. Since photoionization of atoms in the gas play an important role in radiative cooling and heating, we choose to represent the radiation field via the specific photoionization rates $Q_j$ defined as:
\begin{equation}
    Q_j = \frac{c}{n_H}\int_0^\infty \sigma_j(\nu)n_\nu \, d\nu,
    \label{eq:q_j_def}
\end{equation}
where $n_\nu$ is the number density of photons with frequency $\nu$ \citep{gnedin_hollon12}. We explore 58 different such $Q_j$, including rates for all the ionization states of magnesium ($\mathrm{MgI-MgXII}$) and iron ($\mathrm{FeI-FeXXVI}$). We also include the photodissociation rate of molecular hydrogen $Q_\mathrm{LW}$ in order to sample the radiation field at UV energies below the threshold for hydrogen ionization. Since the subsequent ionization states of chemical elements usually have increasing ionization thresholds, ionization rates of all ionization states of a given element sample a wide range of photon energies and hence may serve as good proxies of a radiation field spectrum. These $Q_j$ sample the radiation field across different wavelength ranges, and are determined both by the minimum photon frequency required for the particular ionization and the shape of the ionization cross section $\sigma_j(\nu)$ for higher frequencies. The minimum and maximum values attained given the radiation field parameter ranges from Table~\ref{tab:training_data} are given for rates of particular interest for our analysis in Table~\ref{tab:p_rates}. 

\begin{table}
	\centering
    \begin{tabular}{lllllll}
        \hline
        & HI    & HeI   & CVI   & CaXX   & CI   & NaXI   \\ \hline
        Min & -6.91 & -5.55 & -9.02 & -13.20 & 0.68 & -11.12 \\
        Max & 0.48  & 0.72  & -1.06 & -2.11  & 1.29 & -1.59 \\ \hline
    \end{tabular}
    \caption{Minimum and maximum values of $\log(Q_j/Q_\mathrm{LW})$ for some rates of interest attained in the training data table described in Table~\ref{tab:training_data}.}\label{tab:p_rates}.
\end{table}

\subsection{XGBoost} \label{method:XGBoost}

For this work, we use the gradient-boosted tree algorithm XGBoost.  XGBoost uses an ensemble of trees, where the trees are trained sequentially to predict the residual between the true value and the prediction of the previous trees \citep{chen_guestrin16}. XGBoost is known to perform very well, including outperforming deep learning models, on tabular training data \citep{shwartz-ziv_armon21, grinztajn22}.  Given the tabular setup of our training data as described in Section~\ref{method:data}, XGBoost is an appropriate choice of machine learning model. XGBoost is also well-suited to the high-dimensional input data described in section~\ref{method:data} because the structure of regression trees means that the splitting at any node of any tree is determined by the value of only one feature.  Furthermore, XGBoost includes a hyperparameter which limits the fraction of the input features used for each tree \citep{chen_guestrin16}. An additional reason for the choice of XGBoost over other machine learning models is the availability of tools to compute feature importances, as discussed in section~\ref{method:shap} below. We train XGBoost regression models on a GPU using the tree method \texttt{gpu\_hist}.  Explanations of the XGBoost hyperparameters we varied from their default values can be found in Appendix~\ref{app:hyperparams}.

\subsection{Training Data Preparation} \label{method:data_prep}

The true cooling and heating functions $\Lambda$ and $\Gamma$ vary over several orders of magnitude on the training data, so we train our XGBoost models to predict $\log{\Lambda}$ and $\log{\Gamma}$ as functions of the gas temperature $T$, density $n_H$, metallicity $Z$, and photoionization rates $Q_j$.

Since we only have very sparse sampling in $Z/Z_\odot$, we do not use metallicity as an input feature for our XGBoost models.  Instead, we train separate models at each of the 5 metallicity values in 
Table~\ref{tab:training_data}, and interpolate between model predictions (with the input features fixed) at intermediate metallicities.  This interpolation is discussed further in Section~\ref{method:Z_interp}. We explored incorporating the metallicity $Z$ as an additional XGBoost feature, but found that we could predict cooling and heating function more accurately by combining fixed-metallicity XGBoost models with interpolation in metallicity. This is somewhat unsurprising, since XGBoost models perform a piecewise constant interpolation in each input feature \citep{chen_guestrin16}.  With only 5 data points in metallicity, we can outperform this by choosing a manual fitting function with some knowledge of the expected behavior.

We utilize gas temperature $\log{(T / \mathrm{K})}$ and density $\log{(n_H/\mathrm{cm}^{-3})}$ as input features for all of our XGBoost models.  To describe the radiation field $J_\nu$, we use photoionization rates $Q_j$ to capture how the incident radiation field impacts various elements in the gas. As suggested by the structure of the interpolation table in \citet{gnedin_hollon12}, we expect to need at least 4 distinct $Q_j$ to adequately describe the radiation field dependence. However, all rates $Q_j$ scale linearly with the radiation field amplitude $J_0$.  Hence, to keep our features as uncorrelated as possible, we use the photodissociation rate of molecular hydrogen $Q_\mathrm{LW}$ as a reference rate.  That is, we use $Q_\mathrm{LW}$ as an input to all of our XGBoost models, and scale all other photoionization rates as $Q_j/Q_\mathrm{LW}$ to remove the overall amplitude.  The radiation field features we use are $\log{(Q_\mathrm{LW}/\mathrm{cm}^3 \, \mathrm{s}^{-1})}$ and various sets of $\log{(Q_j/Q_\mathrm{LW})}$ for $j \neq \mathrm{LW}$. 

\begin{figure*}
    \centering
	\includegraphics[width=\textwidth]{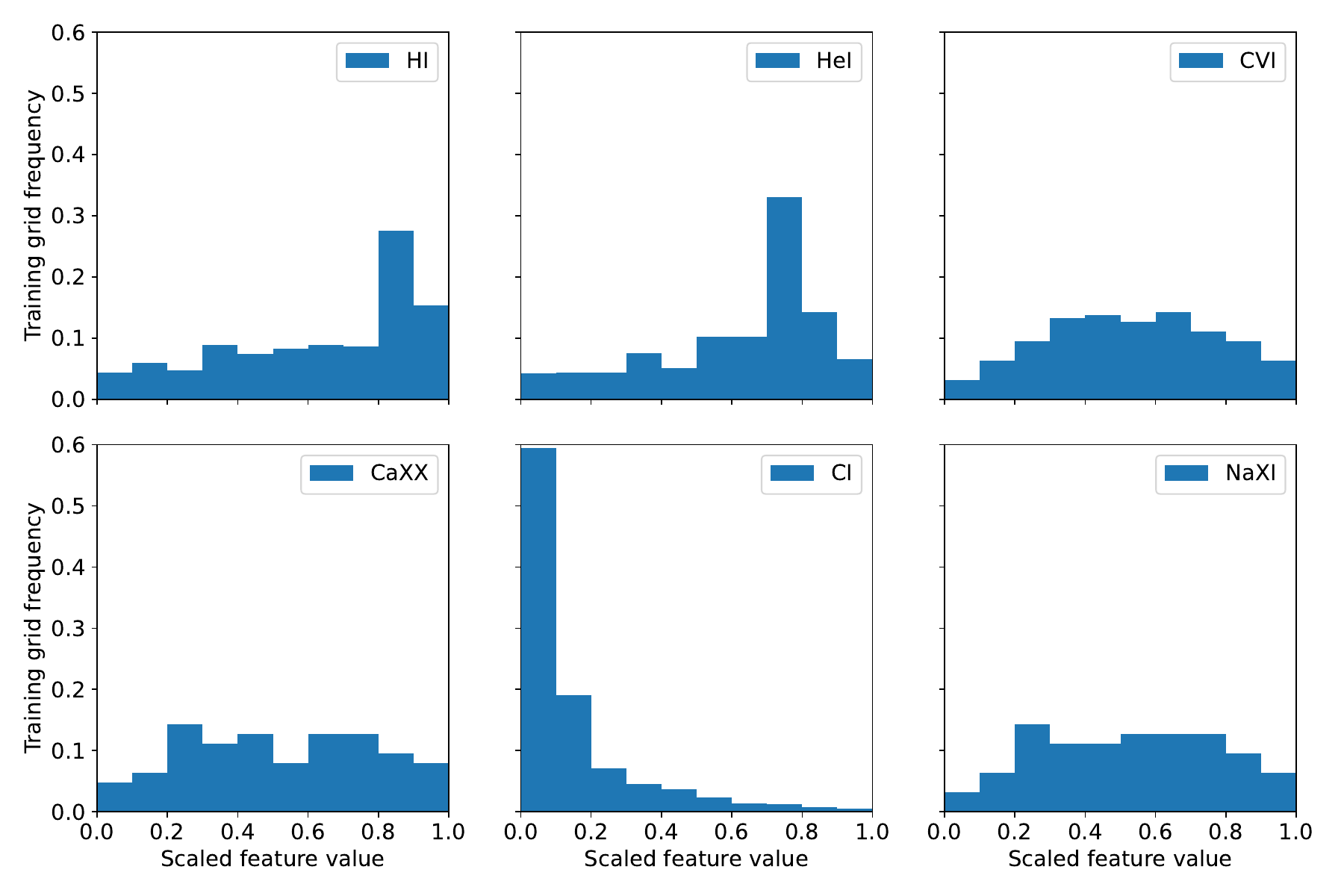}
    \caption{Feature distributions of six scaled rates $\log(Q_j/Q_\mathrm{LW})$.  All have been linearly rescaled to the range $[0,1]$, and their distributions normalized.}
    \label{fig:feat_dist}
\end{figure*}

As seen in Table~\ref{tab:training_data} and Table~\ref{tab:p_rates}, these features have large, and very different, ranges.  For example $1 \leq \log{(T/\mathrm{K})} \leq 9$ and $-6 \leq \log{(n_H/\mathrm{cm}^{-3})} \leq 6$.  To avoid this, we linearly rescale all features to be between 0 and 1 for the training data, using \texttt{MinMaxScaler()} from the \texttt{preprocessing} module of \texttt{scikit-learn} \citep{scikit-learn}. The resulting distributions for 6 $\log{(Q_j/Q_\mathrm{LW})}$ features are shown in Fig.~\ref{fig:feat_dist}.

\subsection{Principal Component Analysis of Rates}
\label{method:pca_rates}

The 58 photoionization rates $Q_j$ we consider have similar ionization thresholds and frequency dependence for their absorption cross-section,  $\sigma_j(\nu)$.  This means that many of the wavelength ranges sampled overlap with each other.  As such, we should expect many of the $Q_j$ to be highly correlated with each other.  As discussed in Section~\ref{method:data_prep}, we always use $\log{(Q_\mathrm{LW}/\mathrm{cm}^3 \, \mathrm{s}^{-1})}$ as a feature to encode the overall radiation field amplitude $J_0$, and consider the scaled rates $\log{(Q_j/Q_\mathrm{LW})}$ for $j \neq \mathrm{LW}$ as candidate features.  There are still 57 of these scaled rates, with highly overlapping wavelength ranges and similar absorption cross sections. An example of this can be seen visually in the right-hand column of Fig.~\ref{fig:feat_dist}, where the feature distributions of $\log{(Q_j/Q_\mathrm{LW})}$ for $j = \mathrm{CVI}$ and $\mathrm{NaXI}$ appear very similar by eye. 

\begin{figure}
    \centering
	\includegraphics[width=\columnwidth]{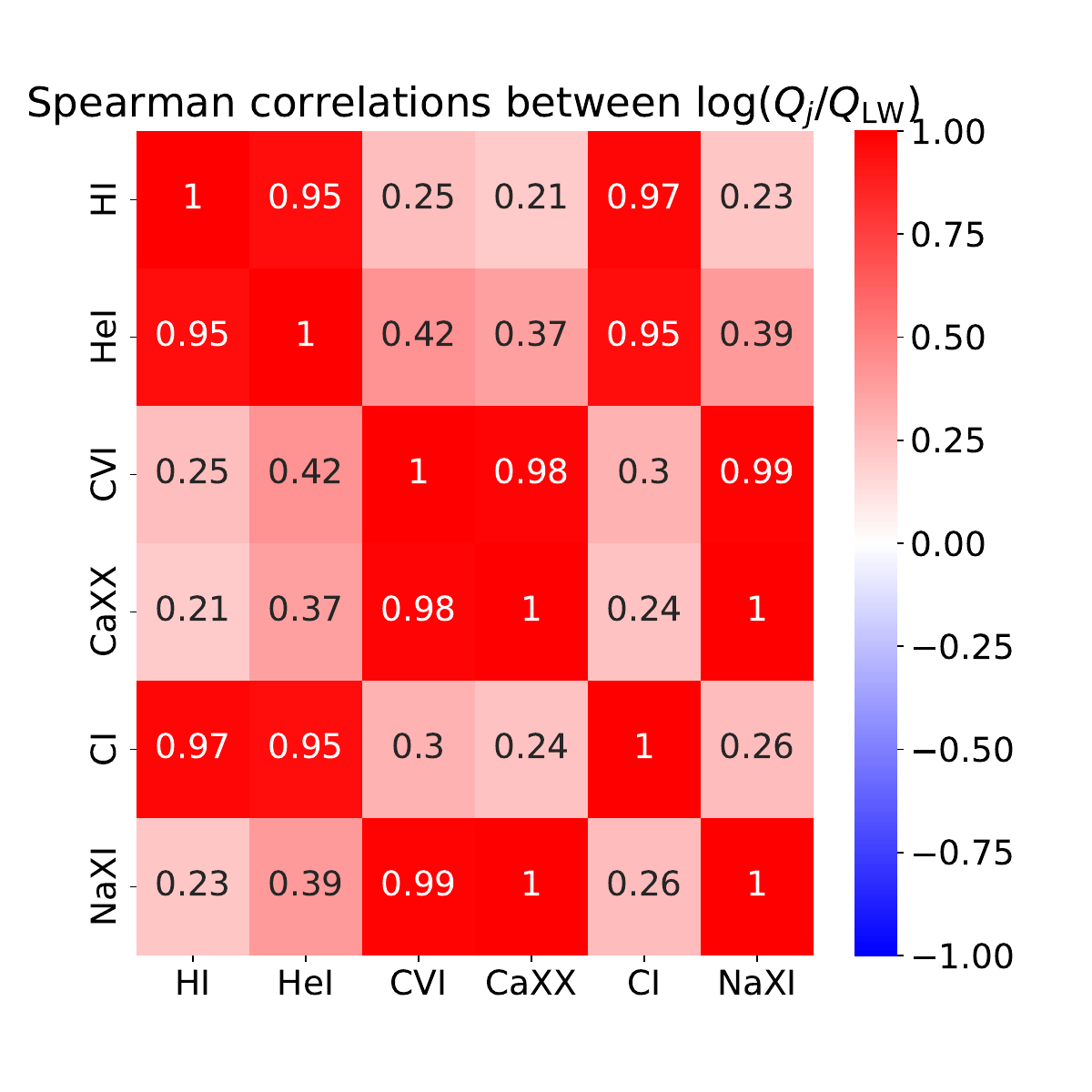}
    \caption{Spearman correlation matrix of the same six scaled rate features $\log{(Q_j/Q_\mathrm{LW})}$ as Fig.~\ref{fig:feat_dist}.}
    \label{fig:feat_corr}
\end{figure}

In order to measure how strongly correlated the photoionization rate features are, we evaluate the correlation matrix of $\log{(Q_j/Q_\mathrm{LW})}, j \neq \mathrm{LW}$ for the data described in Section~\ref{method:data}.  We use the Spearman correlation coefficient, so that the correlations would be the same regardless of whether or not we take the logarithm of the scaled rates (since the logarithm is a monotonically increasing function).  The correlation matrix (for the same scaled rate features $\log{(Q_j/Q_\mathrm{LW})}$ as Fig.~\ref{fig:feat_dist}) is shown in Fig.~\ref{fig:feat_corr}.  Note the high Spearman correlations of $0.95$ or above for several pairs of features, and that the Spearman correlation is positive for all pairs.

To limit both the number of photoionization rate features we need to include and the correlations between them, we utilize tools from principal component analysis (PCA).  In particular, we determine the eigenvalues $\lambda_i$ and normalized eigenvectors $\vec{v}_i$ of the $57 \times 57$ Spearman correlation matrix described above. We find that the principal components of this correlation matrix depend non-trivially on almost all of these rates.  That is, the correlation matrix is close to degenerate, and the principal components are poorly constrained.  For this reason, we depart from PCA and instead rank each $\log{(Q_j/Q_\mathrm{LW})}$ by a "relative importance" $c_j$ defined as the sum of the magnitude of the $j$-th component of each eigenvector $\vec{v}_i(j)$, weighted by the corresponding eigenvalue:
\begin{equation} 
    \label{eq:rate_pca}
    c_j = \sum_i \lambda_i\left|\vec{v}_i(j)\right|.
\end{equation}

As defined by equation~(\ref{eq:rate_pca}), the relative importance for a given scaled rate is essentially a weighted average of the contribution of the scaled rate to each principal component of the correlation matrix, where the weights are the eigenvalue of each principal component.  However, since we do not normalize the sum in equation~(\ref{eq:rate_pca}), only comparisons between different $c_j$ are physically meaningful, and not the magnitude of any individual $c_j$.  In particular, a scaled rate with a larger relative importance $c_j$, on average, contributes more to more important principal components of the correlation matrix.  Limiting the scaled rates we use as features for our XGBoost models to those with sufficiently large relative importance actually limits the number of scaled rates utilized in our analysis.  In contrast, even selecting only the principal component with the largest eigenvalue would technically require the use of all 57 scaled rates, due to the near-degeneracy of the correlation matrix.

\begin{figure*}
    \centering
    \includegraphics[width=\textwidth]{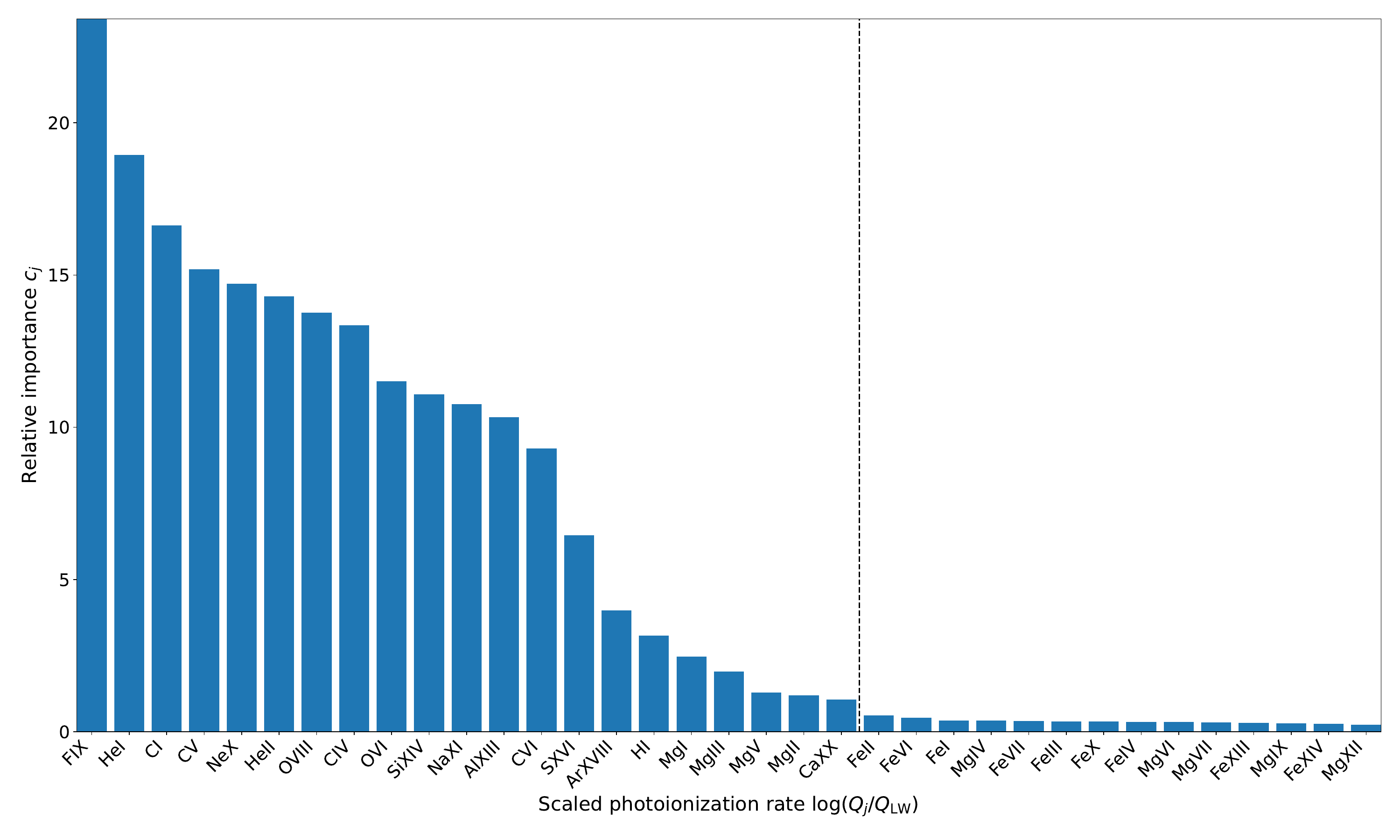}
    \caption{The relative importance $c_j$ defined by equation~(\ref{eq:rate_pca}) for scaled rates $\log{(Q_j/Q_\mathrm{LW})}$, in decreasing order. For simplicity, only the 35 scaled rates with the highest relative importance are shown. We use the 21 rates with the highest significance (those to the left of the vertical dashed line) in our analysis, as all other rates have similarly small significance.}
    \label{fig:eig_sum}
\end{figure*}

We show the relative importance, $c_j$, in Fig.~\ref{fig:eig_sum}.  We choose to select all rates with relative importance $c_j$ larger than that of $j = \mathrm{FeII}$ (i.e. left of the dashed line in Fig.~\ref{fig:eig_sum}), using a natural break in the distribution of $c_j$. Note that all the photoionization rates for iron, and most magnesium rates (with the exception of $\mathrm{MgI-III}$ and $\mathrm{MgV}$) have very low relative importance $c_j$ and are excluded from the remainder of our analysis.  

This leaves $21$ scaled rates, which we include as features in a set of XGBoost models at each metallicity.  We perform a feature importance analysis, discussed in Section~\ref{method:shap}, on these models to guide further downselection of the set of scaled rate features.  We use these models as a baseline comparison for models incorporating fewer scaled rate features.

\subsection{How to Define an Error?}
To train and quantitatively evaluate the performance of our XGBoost models, we must define the error in a predicted cooling or heating function.  We generally define this as:
\begin{equation} 
    \Delta \log \mathcal{F}=\left| (\log{\mathcal{F}})_\mathrm{pred}-(\log{\mathcal{F}})_\mathrm{true} \right|,
\label{eq:error_def}
\end{equation}
where $\mathcal{F}=\Lambda$ or $\Gamma$.

In order to train our XGBoost models, we must choose a `loss function' to minimize on the training set (which is described below in section~\ref{method:model_training}).  This loss function compares the XGBoost predictions to the true values in the data table described in Section~\ref{method:data}, averaged over a set of predictions.  For this purpose, we use the mean squared error (MSE):
\begin{equation}
    \mathrm{MSE} = \langle (\Delta \log \mathcal{F})^2 \rangle,
    \label{eq:MSE_def}
\end{equation}
where the average is over all points in the specific data set in question.

However, we must also keep in mind that for any set of model predictions we will have a \textit{distribution} of errors $\Delta \log \mathcal{F}$, which cannot be completely characterized by the MSE in general.  Instead, we can consider the \textit{cumulative error distribution function} $P(>\Delta \log \mathcal{F})$, which is defined by:
\begin{equation}
    P(>\Delta \log \mathcal{F}) = \frac{\mathrm{Number \, of \, points \, with \, error \, above \,} \Delta \log \mathcal{F}}{\mathrm{Total \, number \, of \, points}}.
    \label{eq:cdf_def}
\end{equation}
In particular, the cumulative error distribution function defined in equation~(\ref{eq:cdf_def}) may include very large ``catastrophic errors''.  While it may have a minimal effect on the overall MSE, minimizing both the frequency and magnitude of these catastrophic errors is crucial for robust prediction of cooling and heating functions.

\subsection{Model Training} \label{method:model_training}

For each set of features, we must perform model validation (hyperparameter optimization), training, and testing for both the cooling function and heating function at $Z/Z_\odot=0, 0.1, 0.3, 1, 3$. That is, we validate, train, and test 10 different XGBoost models for each set of features.

After calculating the photoionization rates, evaluating the cooling or heating function at the desired metallicity, and scaling the features as described above, we split the data into validation, training, and test sets.  Note that we scale the features \textit{before} splitting, so that the minimum and maximum values used to scale each feature could be in any of the testing, training, or validation data. To split the data, we use \texttt{train\_test\_split()} from the \texttt{model\_selection} module of \texttt{scikit-learn}.  We split $20\%$ of the data as the testing set.  Of the remaining $80\%$ of the data, we use $10\%$ for hyperparameter validation (i.e. $8\%$ of the total data).  Our hyperparameter tuning procedure is described in Appendix~\ref{app:hyperparams}.  The remaining $72\%$ of the data is used for model training using hyperparameters obtained from the validation step. These fractions are chosen so that the set used for model training contains the majority of the data, the testing set is large enough to be reasonably qualitatively representative of the feature distributions in the entire data set, and the separation of the validation set does not significantly reduce the amount of data available for model training.  We do not expect small numerical changes to these fractions to significantly affect our results, as long as they satisfy these qualitative constraints. We use the same train-test-validation split for all models.

Using a suitable set of hyperparameters found as described in Appendix~\ref{app:hyperparams}, we then retrain XGBoost models on the $80\%$ training set (but without splitting off any of this data for hyperparameter validation).  We also retrain these XGBoost models on the entire grid for interpolating in $Z$, where we have independent testing data at intermediate values of $Z$.  

\subsection{Feature Importances with SHAP Values} \label{method:shap}

To analyze which scaled photoionization rates have the most impact on the cooling and heating functions, we calculate SHAP values \citep{lundberg_lee17, lundberg18, lundberg20}, using models with the 21 scaled rate features described in Section~\ref{method:pca_rates} and retrained with optimal hyperparameters using 80\% of the training data table, as described in Section~\ref{method:model_training}. We evaluate SHAP values for all features on 500 randomly sampled points from the 20\% test set using the Python package \texttt{shap} \citep{lundberg_lee17, lundberg20}. 

The formula for computing the exact SHAP value for a particular model prediction and input feature comes from game theory, and describes the difference between the true model prediction and what would be expected without including the feature in question.  These feature importances have several desirable properties, including exactly matching the model prediction in question, and always giving a value of $0$ for a feature that does not affect the model prediction.  However, the exact computation involves evaluating machine learning models analogous to the one we wish to explain \textit{using every possible subset of features}. Since this is computationally unfeasible for complex models with many features, SHAP values must often be approximated in practice \citep{lundberg_lee17}.  The specific approximation for tree-based models implemented in the \texttt{shap} package which we use here is described in \citet{lundberg20}.
 
\begin{figure*}
    \centering
    \includegraphics[width = \textwidth]{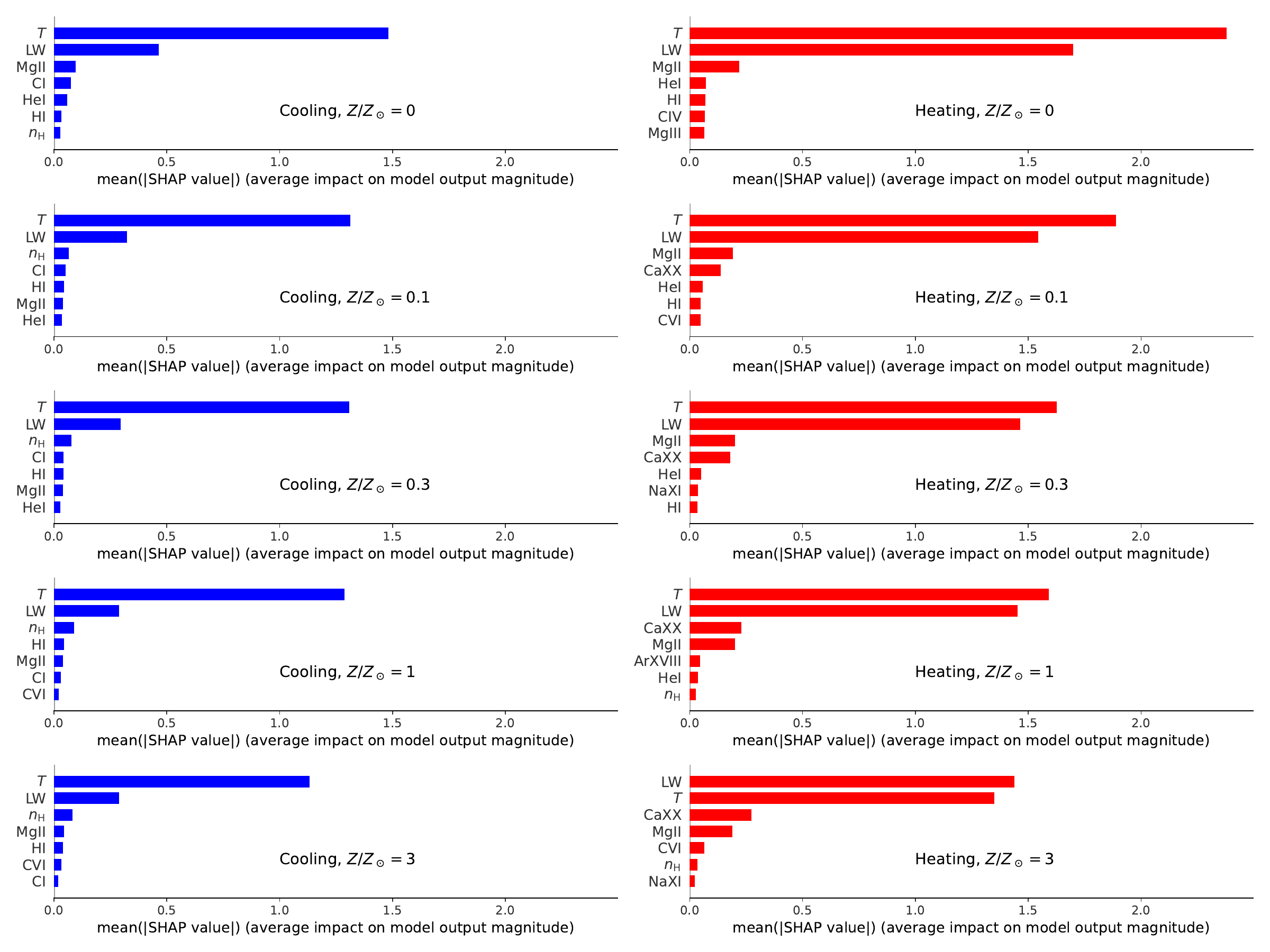}
    \caption{Mean absolute SHAP value for 500 randomly selected points in 20\% of the training grid withheld from model training (see Section~\ref{method:data_prep}) for models trained with 21 scaled rate features selected from the PCA analysis described in Section~\ref{method:pca_rates}.  Cooling function models are shown in blue in the left column, with heating function models in red in the right column.  The rows are for models at the 5 metallicity values in the training data.  Only the 7 features with the largest mean absolute SHAP value are shown.  Note that the full descriptions of the features are $\log(T/\mathrm{K})$ for $T$, $\log(n_\mathrm{H}/\mathrm{cm}^{-3})$ for $n_\mathrm{H}$, $\log(Q_\mathrm{LW}/\mathrm{cm}^3 \, \mathrm{s}^{-1}))$ for $\mathrm{LW}$, and $\log(Q_j/Q_\mathrm{LW})$ otherwise.}
    \label{fig:shap_bar_plot}
\end{figure*}

To compare the importance of the various features, we evaluate the mean absolute SHAP value (since SHAP values can be positive or negative) across the 500 random test points. Ranking these (in decreasing order) gives a description of the most important features (for a given XGBoost model predicting cooling or heating, at a given metallicity). The 7 most important features by this ranking are shown for all models trained using the 21 rates from the PCA described in section~\ref{method:pca_rates} in Fig.~\ref{fig:shap_bar_plot}. For all these models, temperature $\log{(T/\mathrm{K})}$ and radiation field amplitude (described by $\log{(Q_\mathrm{LW}/\mathrm{cm}^3 \, \mathrm{s}^{-1})}$) are the two most important features.  For the cooling function at metallicities $Z/Z_\odot \geq 0.1$, density $\log{(n_H/\mathrm{cm}^{-3})}$ is the third-most important feature by SHAP value, but it is less important at $Z/Z_\odot = 0$ and for the heating function at all metallicities.  The three most important scaled photoionization rates, along with their SHAP value importance rank, are shown in Table~\ref{tab:shap_CF} for the cooling function and Table~\ref{tab:shap_HF} for the heating function.  Several scaled rates ($\mathrm{MgII, CI, HI, HeI}$) have high SHAP values at all metallicities for both cooling and heating, while $\mathrm{CaXX}$ has high SHAP values for heating at all metallicities except for $Z/Z_\odot = 0$.  At the highest metallicities ($Z/Z_\odot = 1, 3$), the rate $\mathrm{CVI}$ reaches high SHAP values. Since the models in Fig.~\ref{fig:shap_bar_plot} are trained at fixed metallicity values $Z/Z_\odot = \{0, 0.1, 0.3, 1, 3\}$, these trends hold only in the metallicity range $0 \leq Z/Z_\odot \leq 3$, and should not be extrapolated beyond this range.  Within this metallicity range, we also do not determine the functional form of the dependence of SHAP values on metallicity, and instead describe only \textit{trends} of increasing or decreasing SHAP values with metallicity.

\begin{table}
	\centering
	\begin{tabular}{lllll} 
		\hline
		$Z/Z_\odot = 10^{-4}$ &  $Z/Z_\odot = 0.1$ &
        $Z/Z_\odot = 0.3$ &
        $Z/Z_\odot = 1$ & 
        $Z/Z_\odot = 3$ \\
		\hline
        MgII (3) & CI (4) & CI (4) & HI (4) & MgII (4) \\
        CI (4) & HI (5) & HI (5) & MgII (5) & HI (5) \\
        HeI (5) & MgII (6) & MgII (6) & CI (6) & CVI (6) \\
		\hline
	\end{tabular}

    \caption{The three most important scaled rates, by mean absolute SHAP value, for 500 randomly sampled test points on cooling function models with the 21 scaled rate features from Section~\ref{method:pca_rates}, retrained on 80\% of the training data table using the optimized hyperparameters found via the procedure in Section~\ref{method:model_training}.}
    \label{tab:shap_CF}
\end{table}

\begin{table}
	\centering
	\begin{tabular}{lllll} 
		\hline
		$Z/Z_\odot = 10^{-4}$ &  $Z/Z_\odot = 0.1$ &
        $Z/Z_\odot = 0.3$ &
        $Z/Z_\odot = 1$ & 
        $Z/Z_\odot = 3$ \\
		\hline
        MgII (3) & MgII (3) & MgII (3) & CaXX (3) & CaXX (3) \\
        HeI (4) & CaXX (4) & CaXX (4) & MgII (4) & MgII (4) \\
        HI (5) & HeI (5) & HeI (5) & ArXVIII (5) & CVI (5) \\
		\hline
	\end{tabular}

    \caption{Same as Table~\ref{tab:shap_HF}, but now for the heating function.}
    \label{tab:shap_HF}
\end{table}

To limit the data size of the XGBoost models, and to compare with the 4 photoionization rates used in the interpolation table in \citet{gnedin_hollon12} ($Q_\mathrm{LW}, Q_\mathrm{HI}, Q_\mathrm{HeI}, Q_\mathrm{CVI}$), we choose to use the 3 scaled rates with the highest SHAP values to train new models (as above, the feature $\log{(Q_\mathrm{LW}/\mathrm{cm}^3 \, \mathrm{s}^{-1})}$ is always used). For cooling and heating at each metallicity, we train models using the 3 most important scaled rate features from our SHAP value analysis in Fig.~\ref{fig:shap_bar_plot}, Table~\ref{tab:shap_CF}, and Table~\ref{tab:shap_HF} using the procedure described in Section~\ref{method:model_training}. We also select several candidate sets of 3 scaled rates that appear frequently in Table~\ref{tab:shap_CF}, Table~\ref{tab:shap_HF}, or both.

\subsection{Interpolation in Metallicity} \label{method:Z_interp}

To approximate $\Lambda$ and $\Gamma$ at values of $Z$ besides the sample points, we must interpolate between the predictions of the 5 fixed-$Z$ models evaluated at the relevant feature values.  Following \citet{gnedin_hollon12}, we perform a quadratic fit to $\Lambda$ and $\Gamma$ at the five metallicity values $Z/Z_\odot=\{0, 0.1, 0.3, 1, 3\}$. However, we found that their quadratic fit sometimes yields unphysical \textit{negative} predictions for $\Lambda$ or $\Gamma$.  

For this reason, we trained our XGBoost models to predict $\Lambda$ and $\Gamma$ at the five metallicities for which we have cooling and heating functions calculated with Cloudy, and perform a quadratic fit between XGBoost predictions that is constrained so as to never predict negative values on the metallicity domain of interest, i.e. $0 \leq Z/Z_\odot \leq 3$.  Specifically, we fit a quadratic in metallicity:
\begin{equation}
    \label{eq:quad_fit}
    f(Z | A, B, C) = A + B \left( \frac{Z}{Z_\odot} \right) + C \left( \frac{Z} {Z_\odot} \right)^2.
\end{equation}
We find the best fit values of the parameters $(A,B,C)$ by numerically minimizing the loss function:
\begin{equation}
    L(A, B, C) = \chi^2(A, B, C) + g(A, B, C).
    \label{eq:quad_fit_loss}
\end{equation}
The $\chi^2$ term is the error between $\mathcal{F}$ ($\Gamma$ or $\Lambda$) predicted by XGBoost, and the quadratic fit, calculated as:
\begin{equation}
    \chi^2(A, B, C) = \sum_i \left[ \mathcal{F}_\mathrm{pred}(Z_i) - f(Z_i | A, B, C) \right]^2,
    \label{eq:chi_squared}
\end{equation}
where the sum is over $Z_i/Z_\odot = \{0, 0.1, 0.3, 1, 3\}$.  The constraint function $g(A, B, C)$ is defined by:
\begin{equation}
    g(A, B, C) = 
    \begin{cases}
         & f(Z | A, B, C) > \min_i \mathcal{F}_\mathrm{pred}(Z_i) \, \mathrm{and} \\
         0  & f(Z | A, B, C) < \max_i \mathcal{F}_\mathrm{pred}(Z_i) \, \forall \, i \\
           & \mathrm{and} 
           \, \forall \, 0 \leq Z/Z_\odot \leq 3, \\
        10^{10} & \mathrm{otherwise},
    \end{cases}
    \label{eq:constraint_func}
\end{equation}
where again $Z_i/Z_\odot = \{0, 0.1, 0.3, 1, 3\}$.
This constraint ensures that the quadratic fit is bounded by the minimum and maximum XGBoost predictions at fixed metallicity.  Our XGBoost models actually predict $\log\mathcal{F}$, so the values of $\mathcal{F}$ predicted by XGBoost are always positive.  Hence, the constraint $g(A, B, C)$ also ensures that the quadratic fit always yields positive values for the metallicity range where we make predictions ($0 \leq Z/Z_\odot \leq 3$).

To evaluate the performance of our XGBoost models combined with a quadratic fit in metallicity, we use a sample of off-grid data, which is randomly sampled in $\log(n_H / \mathrm{cm}^{-3}), \log(J_0 \, \mathrm{cm}^{-3}/n_b/J_\mathrm{MW}), \log{f_Q}, \log{\tau_0},$ and $\alpha$ for the same ranges as in Table~\ref{tab:training_data}, and in the range $[-3, \log{3}]$ for $\log(Z/Z_\odot)$.  We use 8666 such random samples, evaluated at the same $81$ values of  $\log(T/\mathrm{K})$ as in Table~\ref{tab:training_data}, for a total of $701{,}946$ evaluation points.

\section{Results}

\subsection{Comparison on training data}
\label{res:grid_fixed_Z}

We first compare the performance of our XGBoost models at fixed metallicity to the interpolation table in \citet{gnedin_hollon12} on the training data described in Section~\ref{method:data}.  To make this comparison, we use XGBoost models trained on the entirety of this data set, as \citet{gnedin_hollon12} uses that same set to construct their interpolation table.  To understand how accurately the XGBoost models can generalize outside of the training data, we also consider the performance of models trained on only 80\% of the training grid (as described in Section~\ref{method:model_training}) on the 20\% test set withheld from model training. 

To quantify the performance of these models, we consider the cumulative distribution function of errors $P(\Delta >\log \mathcal{F})$ defined in equation~(\ref{eq:cdf_def}).  This is just the frequency of errors \textit{at least as large as} $\Delta \log \mathcal{F}$.  We also consider the mean squared error (MSE, defined in equation~(\ref{eq:MSE_def})).  

At each fixed metallicity, we compare the interpolation table in \citet{gnedin_hollon12} with 3 different XGBoost models: a model trained using the same inputs as \citet{gnedin_hollon12}, i.e. using the scaled photoionization rates $\log(Q_j/Q_\mathrm{LW})$ for $j=\{\mathrm{HI, HeI, CVI}\}$; a model trained using the 21 scaled rates identified from the PCA analysis described in Section~\ref{method:pca_rates}; and a model trained with the top 3 scaled rates from the SHAP value analysis of Section~\ref{method:shap}.  

The approach of \citet{gnedin_hollon12} involves a quadratic fit to Cloudy calculations at the 5 metallicity values in the training data table described in Section~\ref{method:data}, $Z/Z_\odot = \{0, 0.1, 0.3, 1, 3\}$. Here, we focus on $Z/Z_\odot = 0$ and $1$ for comparison with our XGBoost models.

\begin{figure*}
    \centering
    \includegraphics[width = 2\columnwidth]{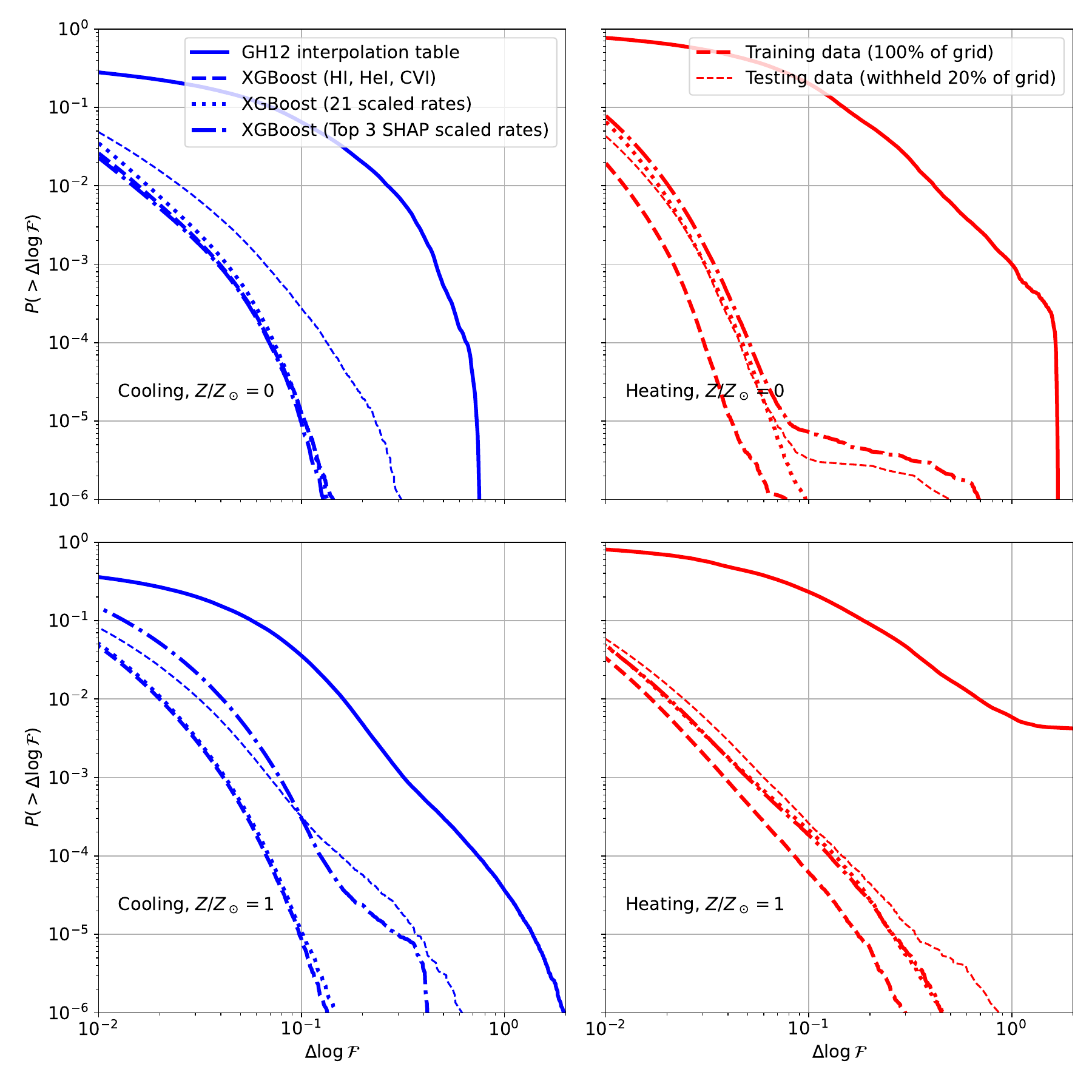}
    \caption{Cumulative distribution function of errors $\Delta \log \mathcal{F}$ (see equation~(\ref{eq:cdf_def})) on the training data (described in Section~\ref{method:data}) for the cooling function $\Lambda$ (left column, in blue) and heating function $\Gamma$ (right column, in red) at fixed metallicity $Z = 0$ (top row) and $Z=Z_\odot$ (bottom row), for the interpolation table in \citet{gnedin_hollon12} (solid lines) and XGBoost models trained using the same scaled photoionization rates $j = \mathrm{HI, HeI, CVI}$ (dashed lines), 21 scaled rates from the PCA analysis described in Section~\ref{method:pca_rates} (dotted lines), and using the top 3 scaled rates from SHAP (dashed-dotted lines, described in Section~\ref{method:shap}.  For the "XGBoost (HI, HeI, CVI)" model, we also include the cumulative error distribution function on 20\% of the training grid withheld from model training for otherwise identical models (thin dashed lines).}
    \label{fig:fixed_Z_cdfs}
\end{figure*}

Fig.~\ref{fig:fixed_Z_cdfs} shows the cumulative distribution function of training errors for the interpolation table in \citet{gnedin_hollon12} and the three types of XGBoost model described above for the cooling and heating function at $Z/Z_\odot = 0$ and $1$ (for the same plots at all metallicities in the training data, see Appendix~\ref{app:all_Z_training_comp}).  The MSEs for all these models are shown for the cooling function in Table~\ref{tab:CF_MSE_fixed_Z} and heating function in Table~\ref{tab:HF_MSE_fixed_Z}.  

\begin{table*}
    \centering
    \begin{tabular}{llllllll}
    \hline
Cooling           &                     & \multicolumn{2}{l}{XGBoost (HI, HeI, CVI)}    & \multicolumn{2}{l}{XGBoost (21 scaled rates)} & \multicolumn{2}{l}{XGBoost (Top 3 SHAP scaled rates)} \\
                  & GH12                  & Training              & Test                  & Training              & Test                  & Training                  & Test                      \\ \hline
$Z/Z_\odot = 0$   & $3.27 \times 10^{-3}$ ($0$) & $1.49 \times 10^{-5}$ & $3.76 \times 10^{-5}$ & $2.11 \times 10^{-5}$ & $4.19 \times 10^{-5}$ & $1.37 \times 10^{-5}$     & $3.60 \times 10^{-5}$     \\
$Z/Z_\odot = 0.1$ & $1.65 \times 10^{-3}$ ($1.63 \times 10^{-4}$)  & $1.36 \times 10^{-5}$ & $3.58 \times 10^{-5}$ & $1.97 \times 10^{-5}$ & $4.33 \times 10^{-5}$ & $2.86 \times 10^{-5}$     & $5.54 \times 10^{-5}$     \\
$Z/Z_\odot = 0.3$ & $1.69 \times 10^{-3}$ ($2.98 \times 10^{-4}$)  & $1.28 \times 10^{-5}$ & $4.04 \times 10^{-5}$ & $1.93 \times 10^{-5}$ & $4.56 \times 10^{-5}$ & $4.39 \times 10^{-5}$     & $8.39 \times 10^{-5}$     \\
$Z/Z_\odot = 1$   & $1.69 \times 10^{-3}$ ($0$) & $2.29 \times 10^{-5}$ & $5.94 \times 10^{-5}$ & $2.51 \times 10^{-5}$ & $6.02 \times 10^{-5}$ & $9.80 \times 10^{-5}$     & $1.74 \times 10^{-4}$     \\
$Z/Z_\odot = 3$   & $1.91 \times 10^{-3}$ ($2.49 
\times 10^{-5}$)  & $3.95 \times 10^{-5}$ & $8.06 \times 10^{-5}$ & $6.83 \times 10^{-5}$ & $1.11 \times 10^{-4}$ & $3.40 \times 10^{-5}$     & $7.84 \times 10^{-5}$     \\ \hline
    \end{tabular}
    \caption{Mean squared errors (MSEs) for the cooling function models shown in Fig.~\ref{fig:fixed_Z_cdfs} and Fig.~\ref{fig:all_z_fixed_z_cdfs}.  For the interpolation table in \citet{gnedin_hollon12}, the MSE is on the entire training grid described in Section~\ref{method:data}. The values in parentheses are the fraction of `catastrophic errors' on the training set where the prediction is negative (so $\Delta \log \mathcal{F} = \infty$).  We remove these points from the MSE calculation.  For the three XGBoost models, the MSE is shown for a model trained and evaluated on the entire training grid ('training' columns), and for a model trained on 80\% of the training grid and evaluated on the 20\% that is withheld (`test' columns).\label{tab:CF_MSE_fixed_Z}}
\end{table*}

\begin{table*}
\centering
\begin{tabular}{llllllll}
\hline
Heating           &                       & \multicolumn{2}{l}{XGBoost (HI, HeI, CVI)}    & \multicolumn{2}{l}{XGBoost (21 scaled rates)} & \multicolumn{2}{l}{XGBoost (Top 3 SHAP scaled rates)} \\
                  & GH12                  & Training              & Test                  & Training              & Test                  & Training                  & Test                      \\ \hline
$Z/Z_\odot = 0$   & $1.27 \times 10^{-2}$ ($0$) & $1.10 \times 10^{-5}$ & $2.08 \times 10^{-5}$ & $3.03 \times 10^{-5}$ & $4.05 \times 10^{-5}$ & $3.41 \times 10^{-5}$     & $5.03 \times 10^{-5}$     \\
$Z/Z_\odot = 0.1$ & $3.26 \times 10^{-2}$ ($0$) & $4.51 \times 10^{-5}$ & $6.74 \times 10^{-5}$ & $2.89 \times 10^{-5}$ & $6.13 \times 10^{-5}$ & $4.45 \times 10^{-5}$     & $8.24 \times 10^{-5}$     \\
$Z/Z_\odot = 0.3$ & $2.19 \times 10^{-2}$ ($0$)& $3.22 \times 10^{-5}$ & $5.79 \times 10^{-5}$ & $3.95 \times 10^{-5}$ & $7.44 \times 10^{-5}$ & $3.09 \times 10^{-5}$     & $6.13 \times 10^{-5}$     \\
$Z/Z_\odot = 1$   & $9.51 \times 10^{-2}$ ($1.84 \times 10^{-3}$) & $1.96 \times 10^{-5}$ & $4.38 \times 10^{-5}$ & $3.28 \times 10^{-5}$ & $6.12 \times 10^{-5}$ & $3.09 \times 10^{-5}$     & $5.75 \times 10^{-5}$     \\
$Z/Z_\odot = 3$   & $2.05 \times 10^{-2}$ ($1.05 \times 10^{-5}$)  & $3.28 \times 10^{-5}$ & $5.82 \times 10^{-5}$ & $1.68 \times 10^{-5}$ & $4.32 \times 10^{-5}$ & $2.16 \times 10^{-5}$     & $4.27 \times 10^{-5}$     \\ \hline
\end{tabular}
\caption{Same as Table~\ref{tab:CF_MSE_fixed_Z}, but now for heating function models.}
\label{tab:HF_MSE_fixed_Z}
\end{table*}

From Fig.~\ref{fig:fixed_Z_cdfs}, Fig.~\ref{fig:all_z_fixed_z_cdfs}, Table~\ref{tab:CF_MSE_fixed_Z}, and Table~\ref{tab:HF_MSE_fixed_Z}, we see that all 3 types of XGBoost model greatly outperform the interpolation table in \citet{gnedin_hollon12} in all cases, with lower MSEs and a smaller cumulative error distribution function for all values shown on the plots ($10^{-2} < \Delta \log \mathcal{F} < 2$).  The MSE on the 20\% of the training table withheld from model training is always larger than that for analogous models trained on the entire table by a factor of order unity.  However, these MSE values are still always smaller than for the interpolation table in \citet{gnedin_hollon12}, and the cumulative error distribution function is similarly lower at the same error values.  

The comparisons between the 3 different types of XGBoost model are not so clear.  In particular, we can see that the cumulative error distribution functions for the XGBoost models (on the entire training table) sometimes intersect in Fig.~\ref{fig:fixed_Z_cdfs} and Fig.~\ref{fig:all_z_fixed_z_cdfs} (see the cooling function at $Z/Z_\odot = 0$ in the upper left of both figures), meaning that we cannot unambiguously identify a single `best-performing' XGBoost model.

\subsection{Comparison on off-grid data}
\label{res:off_grid}

Next, we compare the performance of our XGBoost models combined with the interpolation in metallicity described in Section~\ref{method:Z_interp} to the interpolation table in \citet{gnedin_hollon12} on the sample of off-grid data (also described in Section~\ref{method:Z_interp}).  Note that we are now evaluating the XGBoost models on points not seen in model training, and introducing interpolation between individual XGBoost predictions.  Both of these factors decrease model performance.  Since the rate of change of the energy density of a gas cloud is proportional to the heating function minus the cooling function, as seen equation~(\ref{eq:chf_def}), the physical evolution will usually be dominated by whichever of the cooling or heating function is larger.  Hence, we will also consider the error in the maximum of the cooling and heating function at each point in the evaluation data.

For the interpolation in metallicity, we must select a common set of scaled rate features to use in the XGBoost models at each $Z$ value.  Here, we consider two different sets of 3 scaled rates, $\{\mathrm{CaXX, HI, CVI}\}$ and $\{\mathrm{CI, HI, NaXI}\}$, in addition to the set $\{\mathrm{HI, HeI, CVI}\}$ that makes XGBoost take the same inputs as the interpolation table in \citet{gnedin_hollon12}.  The choice of these sets of scaled rates is motivated by considering rates which have high SHAP value importance across the range of metallicity values, and by the heuristic of needing a photoionization rate with X-ray threshold to capture the quasar part of the ionizing radiation field (parameterized by the power-law slope $\alpha$).  

\begin{figure*}
    \centering
    \includegraphics[width = 2\columnwidth]{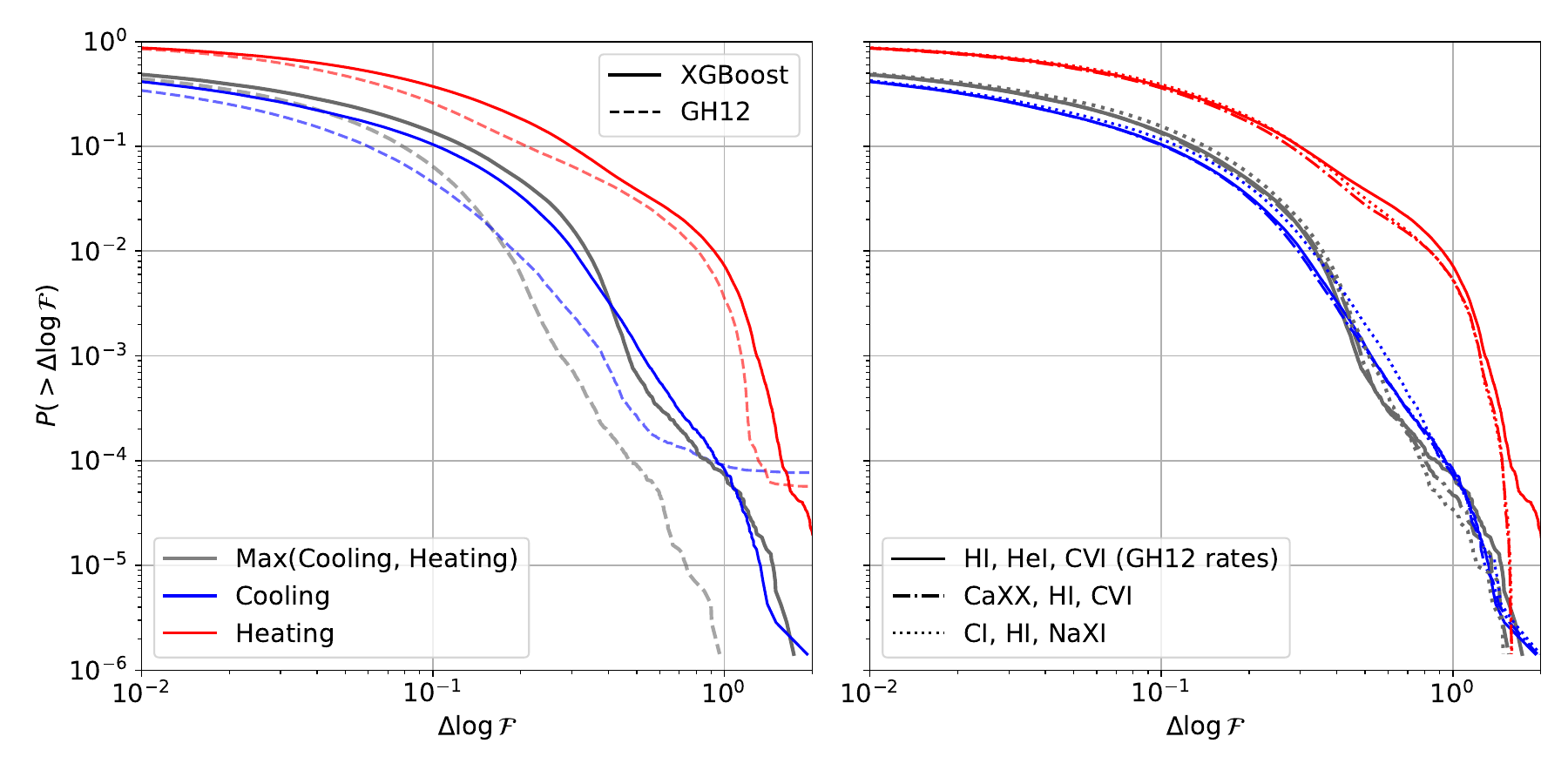}
    \caption{Cumulative distribution function of errors $\Delta \log \mathcal{F}$ on the evaluation data described in Section~\ref{method:Z_interp}.  Left panel: Comparison of results from the interpolation table in \citet{gnedin_hollon12} and the constrained quadratic fit described in Section~\ref{method:Z_interp} between XGBoost models trained with the same scaled rates (HI, HeI, CVI).  Right panel: Comparison of the same constrained quadratic fit between XGBoost models trained with three different sets of scaled rates.}
    \label{fig:eval_data_cdfs}
\end{figure*}

In the left panel of Fig.~\ref{fig:eval_data_cdfs}, we plot the cumulative error distribution function on the off-grid evaluation data described in Section~\ref{method:Z_interp} for the interpolation table in \citet{gnedin_hollon12} and our constrained quadratic interpolation between XGBoost models using the same inputs.  In the right panel, we compare the cumulative error distribution function for our constrained quadratic interpolation between XGBoost models with the three different sets of scaled rates described above.  In Table~\ref{tab:mse_off_grid}, we show the MSEs corresponding to each of the cumulative error distribution functions in Fig.~\ref{fig:eval_data_cdfs}.

\begin{table*}
\begin{tabular}{lllll}
\hline
            & GH12                  & XGBoost (HI, HeI, CVI) & XGBoost (CaXX, HI, CVI) & XGBoost (CI, HI, NaXI) \\ \hline
Max(C, H)   & $2.29 \times 10^{-3}$ ($0$) & $6.89 \times 10^{-3}$  & $6.84 \times 10^{-3}$   & $7.94 \times 10^{-3}$  \\
Cooling (C) & $2.02 \times 10^{-3}$ ($7.55 \times 10^{-5}$)  & $5.26 \times 10^{-3}$  & $5.07 \times 10^{-3}$   & $6.32 \times 10^{-3}$  \\
Heating (H) & $3.00 \times 10^{-2}$ ($5.56 \times 10^{-5}$)  & $4.49 \times 10^{-2}$  & $3.71 \times 10^{-2}$   & $4.06 \times 10^{-2}$ \\ \hline
\end{tabular}
\caption{Mean squared errors on the off-grid data set described in Section~\ref{method:Z_interp} for the models in Fig.~\ref{fig:eval_data_cdfs}. For the GH12 column, the values in parentheses are the fraction of `catastrophic errors' on the evaluation set where the prediction is negative (so $\Delta \log \mathcal{F} = \infty$).  We remove these points from the MSE calculation.}
\label{tab:mse_off_grid}
\end{table*}

From the left panel of Fig.~\ref{fig:eval_data_cdfs}, we see that our XGBoost models plus constrained quadratic interpolation can reduce the frequency of the largest cooling and heating function errors compared to the interpolation table in \citet{gnedin_hollon12}, using the same inputs. From Table~\ref{tab:mse_off_grid}, we see that these XGBoost models (using the scaled rates $\{ \mathrm{HI, HeI, CVI}\}$ yields comparable MSE values to the interpolation table in \citet{gnedin_hollon12}, while (by construction) completely removing the occurrence of `catastrophic errors' for cooling or heating. However, this improvement \textit{does not} affect the prediction of the maximum of the cooling or heating functions. Indeed, both the cumulative error distribution function at each error value shown ($10^{-2} < \Delta \log \mathcal{F} < 2$) and the overall MSE are higher for XGBoost than \citet{gnedin_hollon12}.

The performance of all three XGBoost calculations in the right panel of Fig.~\ref{fig:eval_data_cdfs} are similar.  Neither rate set in this figure produces significant improvement in the performance for the maximum of the cooling or heating function compared to $\{\mathrm{HI, HeI, CVI}\}$, in either MSE or the cumulative error distribution function.  However, the scaled rate set $\{\mathrm{CaXX, HI, CVI}\}$ results in lower MSE, and slightly lower cumulative error distribution in the high-error tail for cooling and heating individually.  Hence, we are able to produce some improvement in overall performance with judicious choices of scaled rate features.

Comparison of Fig.~\ref{fig:fixed_Z_cdfs} and Fig.~\ref{fig:eval_data_cdfs} suggests that, for the training data grid used in this project and \citet{gnedin_hollon12}, there is a bottleneck in model performance on off-grid data due to the interpolation in $Z$.  This is perhaps unsurprising, given that we have only $5$ values of $Z$ to interpolate over, and that we should expect the true dependence of $\Gamma$ and $\Lambda$ on $Z$ to be more complicated than a quadratic.  

\section{Conclusions and Discussion}

In this paper, we train XGBoost models to predict cooling and heating functions at the fixed metallicities $Z/Z_\odot = \{0, 0.1, 0.3, 1, 3\}$, using the same pre-computed Cloudy tables as \citet{gnedin_hollon12}.  We perform a constrained quadratic fit (with the input features held constant) between these XGBoost models to predict cooling and heating functions at arbitrary metallicity.  Using model features selected from a principal component analysis and SHAP value feature importances, we compare both fixed metallicity performance and performance at arbitrary metallicity with results from the interpolation tables in \citet{gnedin_hollon12} in order to assess the potential of XGBoost as an alternative approach for evaluating cooling and heating functions in the presence of generalized radiation fields.  We summarize our findings as follows:

\begin{itemize}
    \item Using the same photoionization rates as \citet{gnedin_hollon12} ($j=\mathrm{LW, HI, HeI, CVI}$) our XGBoost models are able to significantly reduce the frequencies of errors $0.01 < \Delta \log \mathcal{F} < 2$, as well as the mean squared error, at all 5 fixed metallicities. This can be seen in the comparison between solid and dashed thick lines in Fig.~\ref{fig:fixed_Z_cdfs} and Fig.~\ref{fig:all_z_fixed_z_cdfs}. 
    \item Using an eigenvalue and eigenvector analysis of the correlation matrix of scaled photoionization rates, we identified 21 particularly significant scaled photoionization rates (see Fig.~\ref{fig:eig_sum}). 
    \item For models trained using these 21 rates, we identified the most important features using SHAP values (see Fig.~\ref{fig:shap_bar_plot}). Many rates have consistently high importance for both cooling and heating across all five metallicity values, especially $\mathrm{MgII, CI, HI,}$ and $\mathrm{HeI}$.  The rate $\mathrm{CaXX}$ has high importance for the heating function, but not cooling.  The rate $\mathrm{CVI}$ has low importance at low metallicities, but emerges as one of the more important rates for the highest metallicity $Z/Z_\odot = 3$ for both the cooling and heating functions. We expect trends in mean SHAP values with metallicity to be robust for the metallicity range we sample, $0 \leq Z/Z_\odot \leq 3$.  Higher mean SHAP values for a given rate signifies that the wavelengths sampled by that photoionization rate affect cooling or heating more strongly at that metallicity.
    \item The shape of the cumulative error distribution functions for various fixed Z models (see the dashed, dotted, and dot-dashed lines in Fig.~\ref{fig:fixed_Z_cdfs}) cannot be adequately described by a single parameter. 
    While the MSE (see the corresponding values in Table~\ref{tab:CF_MSE_fixed_Z} and Table~\ref{tab:HF_MSE_fixed_Z}) describes the overall normalization of these curves, they also can differ in overall shape, particularly in how steeply the cumulative error distribution function falls off at high errors. 
    This generally prevents determination of a single clearly `best-performing' model.
    \item Using the same scaled rates as \citet{gnedin_hollon12}, we are able to reduce the frequency of the largest errors $\log\mathcal{F} \geq 1$ for cooling or heating individually. We were able to achieve further reduction with other rate sets suggested by a combination of SHAP values and physical heuristics, such as $\{\mathrm{CaXX, HI, CVI}\}$ (see Fig.~\ref{fig:eval_data_cdfs}). 
\end{itemize}

The reduction in the frequency of cooling or heating function errors $\Delta \log \mathcal{F} \geq 1$ that we achieve at arbitrary metallicity (see Fig.~\ref{fig:eval_data_cdfs} and Table~\ref{tab:mse_off_grid}) is much more modest than that seen for our individual fixed metallicity XGBoost models (see Fig.~\ref{fig:fixed_Z_cdfs}, Table~\ref{tab:CF_MSE_fixed_Z}, and Table~\ref{tab:HF_MSE_fixed_Z}).  This is likely due to the small number of samples we have to perform a fit in metallicity (the five $Z/Z_\odot$ values in Table~\ref{tab:training_data}).

Our metallicity fit, seen in equation~(\ref{eq:quad_fit}), assumes that the dependence of cooling and heating functions on metallicity (with all other parameters fixed) is quadratic.  If the only processes involved were two-body collisions, then this assumption would be exactly correct \textit{for the cooling and heating functions computed by Cloudy}. However, the Cloudy training data includes additional processes involving multiple electrons \citep{ferland98}.  Furthermore, even small errors in our XGBoost predictions at fixed metallicity imply that the dependence of these predictions on metallicity will not exactly match the metallicity dependence of the Cloudy calculations they approximate. Also, we introduced an additional unphysical constraint on our quadratic fits in order to prevent negative predicted cooling and heating functions. These all suggest that a different, more complex metallicity fitting function is needed to accurately capture the metallicity dependence.

Hence, this paper suggests future work running Cloudy at intermediate metallicity values between the five used here, in order to implement a more flexible and accurate fit in metallicity and better understand the metallicity dependence of cooling and heating functions.  This additional Cloudy data could also enable further exploration of how the importances of various radiation field parameters changes with metallicity.

\section*{Acknowledgements}
This manuscript has been co-authored by Fermi Research Alliance, LLC under Contract No. DE-AC02-07CH11359 with the U.S. Department of Energy, Office of Science, Office of High Energy Physics.  This research was also supported in part through computational resources and services provided by Advanced Research Computing (ARC), a division of Information and Technology Services (ITS) at the University of Michigan, Ann Arbor, in particular the Great Lakes cluster and the U-M Research Computing Package.

DR acknowledges funding provided by the Leinweber Graduate Fellowship at the University of Michigan; the National Aeronautics and Space Administration (NASA), under award number 80NSSC20M0124, Michigan Space Grant Consortium (MSGC); and the U.S. Department of Energy, Office of Science, Office of Workforce Development for Teachers and Scientists, Office of Science Graduate Student Research (SCGSR) program. The SCGSR program is administered by the Oak Ridge Institute for Science and Education for the DOE under contract number DE‐SC0014664.  CA acknowledges support from the Leinweber Center for Theoretical Physics at the University of Michigan and Department of Energy grant DE-SC009193. The authors would like to thank the anonymous reviewer for their helpful and constructive comments.

\section*{Data Availability}

The XGBoost pipeline is maintained on GitHub at \url{https://github.com/davidbrobins/ml_chf}.  The Cloudy tables used for model training and evaluation, and our trained XGBoost models, can be made available upon request.


\bibliographystyle{mnras}
\bibliography{main} 

\begin{thebibliography}{}
\makeatletter
\relax
\def\mn@urlcharsother{\let\do\@makeother \do\$\do\&\do\#\do\^\do\_\do\%\do\~}
\def\mn@doi{\begingroup\mn@urlcharsother \@ifnextchar [ {\mn@doi@}
  {\mn@doi@[]}}
\def\mn@doi@[#1]#2{\def\@tempa{#1}\ifx\@tempa\@empty \href
  {http://dx.doi.org/#2} {doi:#2}\else \href {http://dx.doi.org/#2} {#1}\fi
  \endgroup}
\def\mn@eprint#1#2{\mn@eprint@#1:#2::\@nil}
\def\mn@eprint@arXiv#1{\href {http://arxiv.org/abs/#1} {{\tt arXiv:#1}}}
\def\mn@eprint@dblp#1{\href {http://dblp.uni-trier.de/rec/bibtex/#1.xml}
  {dblp:#1}}
\def\mn@eprint@#1:#2:#3:#4\@nil{\def\@tempa {#1}\def\@tempb {#2}\def\@tempc
  {#3}\ifx \@tempc \@empty \let \@tempc \@tempb \let \@tempb \@tempa \fi \ifx
  \@tempb \@empty \def\@tempb {arXiv}\fi \@ifundefined
  {mn@eprint@\@tempb}{\@tempb:\@tempc}{\expandafter \expandafter \csname
  mn@eprint@\@tempb\endcsname \expandafter{\@tempc}}}

\bibitem[\protect\citeauthoryear{{Andrae}, {Rix}  \& {Chandra}}{{Andrae}
  et~al.}{2023}]{andrae23}
{Andrae} R.,  {Rix} H.-W.,   {Chandra} V.,  2023, \mn@doi [\apjs]
  {10.3847/1538-4365/acd53e}, \href
  {https://ui.adsabs.harvard.edu/abs/2023ApJS..267....8A} {267, 8}

\bibitem[\protect\citeauthoryear{{Anninos}, {Zhang}, {Abel}  \&
  {Norman}}{{Anninos} et~al.}{1997}]{anninos97}
{Anninos} P.,  {Zhang} Y.,  {Abel} T.,   {Norman} M.~L.,  1997, \mn@doi [\na]
  {10.1016/S1384-1076(97)00009-2}, \href
  {https://ui.adsabs.harvard.edu/abs/1997NewA....2..209A} {2, 209}

\bibitem[\protect\citeauthoryear{{Arnaud} \& {Rothenflug}}{{Arnaud} \&
  {Rothenflug}}{1985}]{arnaud85}
{Arnaud} M.,  {Rothenflug} R.,  1985, \aaps, \href
  {https://ui.adsabs.harvard.edu/abs/1985A&AS...60..425A} {60, 425}

\bibitem[\protect\citeauthoryear{{Baes}, {Dejonghe}  \& {Davies}}{{Baes}
  et~al.}{2005}]{baes05}
{Baes} M.,  {Dejonghe} H.,   {Davies} J.~I.,  2005, in {Popescu} C.~C.,
  {Tuffs} R.~J.,  eds,  American Institute of Physics Conference Series Vol.
  761, The Spectral Energy Distributions of Gas-Rich Galaxies: Confronting
  Models with Data. pp 27--38 (\mn@eprint {arXiv} {astro-ph/0503483}),
  \mn@doi{10.1063/1.1913913}

\bibitem[\protect\citeauthoryear{Bautista \& Kallman}{Bautista \&
  Kallman}{2001}]{bautista01}
Bautista M.~A.,  Kallman T.~R.,  2001, \mn@doi [The Astrophysical Journal
  Supplement Series] {10.1086/320363}, 134, 139

\bibitem[\protect\citeauthoryear{{Benson}}{{Benson}}{2010}]{benson10}
{Benson} A.~J.,  2010, \mn@doi [\physrep] {10.1016/j.physrep.2010.06.001},
  \href {https://ui.adsabs.harvard.edu/abs/2010PhR...495...33B} {495, 33}

\bibitem[\protect\citeauthoryear{{Bertschinger}}{{Bertschinger}}{1985}]{bertschinger85}
{Bertschinger} E.,  1985, \mn@doi [\apjs] {10.1086/191028}, \href
  {https://ui.adsabs.harvard.edu/abs/1985ApJS...58...39B} {58, 39}

\bibitem[\protect\citeauthoryear{{Binney}}{{Binney}}{1977}]{binney77}
{Binney} J.,  1977, \mn@doi [\apj] {10.1086/155378}, \href
  {https://ui.adsabs.harvard.edu/abs/1977ApJ...215..483B} {215, 483}

\bibitem[\protect\citeauthoryear{{Bovino}, {Grassi}, {Capelo}, {Schleicher}  \&
  {Banerjee}}{{Bovino} et~al.}{2016}]{bovino16}
{Bovino} S.,  {Grassi} T.,  {Capelo} P.~R.,  {Schleicher} D.~R.~G.,
  {Banerjee} R.,  2016, \mn@doi [\aap] {10.1051/0004-6361/201628158}, \href
  {https://ui.adsabs.harvard.edu/abs/2016A&A...590A..15B} {590, A15}

\bibitem[\protect\citeauthoryear{{Brooks}, {Governato}, {Quinn}, {Brook}  \&
  {Wadsley}}{{Brooks} et~al.}{2009}]{brooks09}
{Brooks} A.~M.,  {Governato} F.,  {Quinn} T.,  {Brook} C.~B.,   {Wadsley} J.,
  2009, \mn@doi [\apj] {10.1088/0004-637X/694/1/396}, \href
  {https://ui.adsabs.harvard.edu/abs/2009ApJ...694..396B} {694, 396}

\bibitem[\protect\citeauthoryear{{Calderon} \& {Berlind}}{{Calderon} \&
  {Berlind}}{2019}]{calderon_berlind19}
{Calderon} V.~F.,  {Berlind} A.~A.,  2019, \mn@doi [\mnras]
  {10.1093/mnras/stz2775}, \href
  {https://ui.adsabs.harvard.edu/abs/2019MNRAS.490.2367C} {490, 2367}

\bibitem[\protect\citeauthoryear{{Chang}, {Hsieh}, {Wang}, {Lin}, {Lim},
  {Toba}, {Zhong}  \& {Chang}}{{Chang} et~al.}{2021}]{chang21}
{Chang} Y.-Y.,  {Hsieh} B.-C.,  {Wang} W.-H.,  {Lin} Y.-T.,  {Lim} C.-F.,
  {Toba} Y.,  {Zhong} Y.,   {Chang} S.-Y.,  2021, \mn@doi [\apj]
  {10.3847/1538-4357/ac167c}, \href
  {https://ui.adsabs.harvard.edu/abs/2021ApJ...920...68C} {920, 68}

\bibitem[\protect\citeauthoryear{{Chatzikos} et~al.,}{{Chatzikos}
  et~al.}{2023}]{chatzikos23}
{Chatzikos} M.,  et~al., 2023, \mn@doi [arXiv e-prints]
  {10.48550/arXiv.2308.06396}, \href
  {https://ui.adsabs.harvard.edu/abs/2023arXiv230806396C} {p. arXiv:2308.06396}

\bibitem[\protect\citeauthoryear{Chen \& Guestrin}{Chen \&
  Guestrin}{2016}]{chen_guestrin16}
Chen T.,  Guestrin C.,  2016, Proceedings of the 22nd ACM SIGKDD International
  Conference on Knowledge Discovery and Data Mining

\bibitem[\protect\citeauthoryear{{Cole}, {Aragon-Salamanca}, {Frenk}, {Navarro}
   \& {Zepf}}{{Cole} et~al.}{1994}]{cole94}
{Cole} S.,  {Aragon-Salamanca} A.,  {Frenk} C.~S.,  {Navarro} J.~F.,   {Zepf}
  S.~E.,  1994, \mn@doi [\mnras] {10.1093/mnras/271.4.781}, \href
  {https://ui.adsabs.harvard.edu/abs/1994MNRAS.271..781C} {271, 781}

\bibitem[\protect\citeauthoryear{{Cox} \& {Tucker}}{{Cox} \&
  {Tucker}}{1969}]{cox_tucker69}
{Cox} D.~P.,  {Tucker} W.~H.,  1969, \mn@doi [\apj] {10.1086/150144}, \href
  {https://ui.adsabs.harvard.edu/abs/1969ApJ...157.1157C} {157, 1157}

\bibitem[\protect\citeauthoryear{{Croton} et~al.,}{{Croton}
  et~al.}{2006}]{croton06}
{Croton} D.~J.,  et~al., 2006, \mn@doi [\mnras]
  {10.1111/j.1365-2966.2005.09675.x}, \href
  {https://ui.adsabs.harvard.edu/abs/2006MNRAS.365...11C} {365, 11}

\bibitem[\protect\citeauthoryear{{Dalgarno} \& {McCray}}{{Dalgarno} \&
  {McCray}}{1972}]{dalgarno72}
{Dalgarno} A.,  {McCray} R.~A.,  1972, \mn@doi [\araa]
  {10.1146/annurev.aa.10.090172.002111}, \href
  {https://ui.adsabs.harvard.edu/abs/1972ARA&A..10..375D} {10, 375}

\bibitem[\protect\citeauthoryear{Dang, Chen, Li  \& Shu}{Dang
  et~al.}{2022}]{dang22}
Dang Y.,  Chen Z.,  Li H.,   Shu H.,  2022, Applied Artificial Intelligence, 36

\bibitem[\protect\citeauthoryear{{Draine}}{{Draine}}{1978}]{draine78}
{Draine} B.~T.,  1978, \mn@doi [\apjs] {10.1086/190513}, \href
  {https://ui.adsabs.harvard.edu/abs/1978ApJS...36..595D} {36, 595}

\bibitem[\protect\citeauthoryear{{Dumont}, {Abrassart}  \& {Collin}}{{Dumont}
  et~al.}{2000}]{dumont00}
{Dumont} A.~M.,  {Abrassart} A.,   {Collin} S.,  2000, \mn@doi [\aap]
  {10.48550/arXiv.astro-ph/0003220}, \href
  {https://ui.adsabs.harvard.edu/abs/2000A&A...357..823D} {357, 823}

\bibitem[\protect\citeauthoryear{{Dwek}}{{Dwek}}{1998}]{dwek98}
{Dwek} E.,  1998, \mn@doi [\apj] {10.1086/305829}, \href
  {https://ui.adsabs.harvard.edu/abs/1998ApJ...501..643D} {501, 643}

\bibitem[\protect\citeauthoryear{{Ercolano}, {Barlow}, {Storey}  \&
  {Liu}}{{Ercolano} et~al.}{2003}]{ercolano03}
{Ercolano} B.,  {Barlow} M.~J.,  {Storey} P.~J.,   {Liu} X.~W.,  2003, \mn@doi
  [\mnras] {10.1046/j.1365-8711.2003.06371.x}, \href
  {https://ui.adsabs.harvard.edu/abs/2003MNRAS.340.1136E} {340, 1136}

\bibitem[\protect\citeauthoryear{{Fardal}, {Katz}, {Gardner}, {Hernquist},
  {Weinberg}  \& {Dav{\'e}}}{{Fardal} et~al.}{2001}]{fardal01}
{Fardal} M.~A.,  {Katz} N.,  {Gardner} J.~P.,  {Hernquist} L.,  {Weinberg}
  D.~H.,   {Dav{\'e}} R.,  2001, \mn@doi [\apj] {10.1086/323519}, \href
  {https://ui.adsabs.harvard.edu/abs/2001ApJ...562..605F} {562, 605}

\bibitem[\protect\citeauthoryear{{Ferland}}{{Ferland}}{1993}]{ferland93}
{Ferland} G.~J.,  1993, in {Weinberger} R.,  {Acker} A.,  eds,  Vol. 155,
  Planetary Nebulae. p.~123

\bibitem[\protect\citeauthoryear{{Ferland}}{{Ferland}}{2009}]{ferland09}
{Ferland} G.~J.,  2009, \mn@doi [\aap] {10.1051/0004-6361/200912165}, \href
  {https://ui.adsabs.harvard.edu/abs/2009A&A...500..299F} {500, 299}

\bibitem[\protect\citeauthoryear{{Ferland}, {Korista}  \& {Verner}}{{Ferland}
  et~al.}{1997}]{ferland97}
{Ferland} G.~J.,  {Korista} K.~T.,   {Verner} D.~A.,  1997, in {Hunt} G.,
  {Payne} H.,  eds,  Astronomical Society of the Pacific Conference Series Vol.
  125, Astronomical Data Analysis Software and Systems VI. p.~213

\bibitem[\protect\citeauthoryear{{Ferland}, {Korista}, {Verner}, {Ferguson},
  {Kingdon}  \& {Verner}}{{Ferland} et~al.}{1998}]{ferland98}
{Ferland} G.~J.,  {Korista} K.~T.,  {Verner} D.~A.,  {Ferguson} J.~W.,
  {Kingdon} J.~B.,   {Verner} E.~M.,  1998, \mn@doi [\pasp] {10.1086/316190},
  \href {https://ui.adsabs.harvard.edu/abs/1998PASP..110..761F} {110, 761}

\bibitem[\protect\citeauthoryear{{Ferland} et~al.,}{{Ferland}
  et~al.}{2013}]{ferland13}
{Ferland} G.~J.,  et~al., 2013, \mn@doi [\rmxaa] {10.48550/arXiv.1302.4485},
  \href {https://ui.adsabs.harvard.edu/abs/2013RMxAA..49..137F} {49, 137}

\bibitem[\protect\citeauthoryear{{Ferland} et~al.,}{{Ferland}
  et~al.}{2017}]{ferland17}
{Ferland} G.~J.,  et~al., 2017, \mn@doi [\rmxaa] {10.48550/arXiv.1705.10877},
  \href {https://ui.adsabs.harvard.edu/abs/2017RMxAA..53..385F} {53, 385}

\bibitem[\protect\citeauthoryear{{Fu}, {Wu}, {Yang}, {Brown}, {Feng}, {Ma}  \&
  {Li}}{{Fu} et~al.}{2021}]{fu21}
{Fu} Y.,  {Wu} X.-B.,  {Yang} Q.,  {Brown} A. G.~A.,  {Feng} X.,  {Ma} Q.,
  {Li} S.,  2021, \mn@doi [\apjs] {10.3847/1538-4365/abe85e}, \href
  {https://ui.adsabs.harvard.edu/abs/2021ApJS..254....6F} {254, 6}

\bibitem[\protect\citeauthoryear{{Galligan}, {Katz}, {Kimm}, {Rosdahl},
  {Blaizot}, {Devriendt}  \& {Slyz}}{{Galligan} et~al.}{2019}]{galligan19}
{Galligan} T.~P.,  {Katz} H.,  {Kimm} T.,  {Rosdahl} J.,  {Blaizot} J.,
  {Devriendt} J.,   {Slyz} A.,  2019, \mn@doi [arXiv e-prints]
  {10.48550/arXiv.1901.01264}, \href
  {https://ui.adsabs.harvard.edu/abs/2019arXiv190101264G} {p. arXiv:1901.01264}

\bibitem[\protect\citeauthoryear{{Gnat} \& {Sternberg}}{{Gnat} \&
  {Sternberg}}{2007}]{gnat_sternberg07}
{Gnat} O.,  {Sternberg} A.,  2007, \mn@doi [\apjs] {10.1086/509786}, \href
  {https://ui.adsabs.harvard.edu/abs/2007ApJS..168..213G} {168, 213}

\bibitem[\protect\citeauthoryear{{Gnedin} \& {Hollon}}{{Gnedin} \&
  {Hollon}}{2012}]{gnedin_hollon12}
{Gnedin} N.~Y.,  {Hollon} N.,  2012, \mn@doi [\apjs]
  {10.1088/0067-0049/202/2/13}, \href
  {https://ui.adsabs.harvard.edu/abs/2012ApJS..202...13G} {202, 13}

\bibitem[\protect\citeauthoryear{{Golob}, {Sawicki}, {Goulding}  \&
  {Coupon}}{{Golob} et~al.}{2021}]{golob21}
{Golob} A.,  {Sawicki} M.,  {Goulding} A.~D.,   {Coupon} J.,  2021, \mn@doi
  [\mnras] {10.1093/mnras/stab719}, \href
  {https://ui.adsabs.harvard.edu/abs/2021MNRAS.503.4136G} {503, 4136}

\bibitem[\protect\citeauthoryear{{Grassi}, {Krstic}, {Merlin}, {Buonomo},
  {Piovan}  \& {Chiosi}}{{Grassi} et~al.}{2011}]{grassi11}
{Grassi} T.,  {Krstic} P.,  {Merlin} E.,  {Buonomo} U.,  {Piovan} L.,
  {Chiosi} C.,  2011, \mn@doi [\aap] {10.1051/0004-6361/200913779}, \href
  {https://ui.adsabs.harvard.edu/abs/2011A&A...533A.123G} {533, A123}

\bibitem[\protect\citeauthoryear{{Grinsztajn}, {Oyallon}  \&
  {Varoquaux}}{{Grinsztajn} et~al.}{2022}]{grinztajn22}
{Grinsztajn} L.,  {Oyallon} E.,   {Varoquaux} G.,  2022, \mn@doi [arXiv
  e-prints] {10.48550/arXiv.2207.08815}, \href
  {https://ui.adsabs.harvard.edu/abs/2022arXiv220708815G} {p. arXiv:2207.08815}

\bibitem[\protect\citeauthoryear{{Guhathakurta} \& {Draine}}{{Guhathakurta} \&
  {Draine}}{1989}]{guhathakurta_draine89}
{Guhathakurta} P.,  {Draine} B.~T.,  1989, \mn@doi [\apj] {10.1086/167899},
  \href {https://ui.adsabs.harvard.edu/abs/1989ApJ...345..230G} {345, 230}

\bibitem[\protect\citeauthoryear{{Gutcke}, {Pakmor}, {Naab}  \&
  {Springel}}{{Gutcke} et~al.}{2021}]{gutcke21}
{Gutcke} T.~A.,  {Pakmor} R.,  {Naab} T.,   {Springel} V.,  2021, \mn@doi
  [\mnras] {10.1093/mnras/staa3875}, \href
  {https://ui.adsabs.harvard.edu/abs/2021MNRAS.501.5597G} {501, 5597}

\bibitem[\protect\citeauthoryear{{Hayden} et~al.,}{{Hayden}
  et~al.}{2022}]{hayden20}
{Hayden} M.~R.,  et~al., 2022, \mn@doi [\mnras] {10.1093/mnras/stac2787}, \href
  {https://ui.adsabs.harvard.edu/abs/2022MNRAS.517.5325H} {517, 5325}

\bibitem[\protect\citeauthoryear{Head, Kumar, Nahrstaedt, Louppe  \&
  Shcherbatyi}{Head et~al.}{2021}]{scikit-optimize}
Head T.,  Kumar M.,  Nahrstaedt H.,  Louppe G.,   Shcherbatyi I.,  2021,
  scikit-optimize/scikit-optimize, \mn@doi{10.5281/zenodo.5565057}, \url
  {https://doi.org/10.5281/zenodo.5565057}

\bibitem[\protect\citeauthoryear{{Heyl}, {Butterworth}  \& {Viti}}{{Heyl}
  et~al.}{2023}]{heyl23}
{Heyl} J.,  {Butterworth} J.,   {Viti} S.,  2023, \mn@doi [\mnras]
  {10.1093/mnras/stad2814}, \href
  {https://ui.adsabs.harvard.edu/abs/2023MNRAS.526..404H} {526, 404}

\bibitem[\protect\citeauthoryear{{Hopkins}, {Quataert}  \& {Murray}}{{Hopkins}
  et~al.}{2011}]{hopkins11}
{Hopkins} P.~F.,  {Quataert} E.,   {Murray} N.,  2011, \mn@doi [\mnras]
  {10.1111/j.1365-2966.2011.19306.x}, \href
  {https://ui.adsabs.harvard.edu/abs/2011MNRAS.417..950H} {417, 950}

\bibitem[\protect\citeauthoryear{{Hughes}, {Bailer-Jones}  \& {Jamal}}{{Hughes}
  et~al.}{2022}]{hughes22}
{Hughes} A. C.~N.,  {Bailer-Jones} C. A.~L.,   {Jamal} S.,  2022, \mn@doi
  [\aap] {10.1051/0004-6361/202244859}, \href
  {https://ui.adsabs.harvard.edu/abs/2022A&A...668A..99H} {668, A99}

\bibitem[\protect\citeauthoryear{{Ivanov}, {Tsizh}, {Ullmann}, {Panos}  \&
  {Voloshynovskiy}}{{Ivanov} et~al.}{2021}]{ivanov21}
{Ivanov} S.,  {Tsizh} M.,  {Ullmann} D.,  {Panos} B.,   {Voloshynovskiy} S.,
  2021, \mn@doi [Astronomy and Computing] {10.1016/j.ascom.2021.100473}, \href
  {https://ui.adsabs.harvard.edu/abs/2021A&C....3600473I} {36, 100473}

\bibitem[\protect\citeauthoryear{Jia, Sun, Lian  \& Hou}{Jia
  et~al.}{2022}]{jia22}
Jia W.,  Sun M.,  Lian J.,   Hou S.,  2022, \mn@doi [Complex & Intelligent
  Systems] {10.1007/s40747-021-00637-x}, 8

\bibitem[\protect\citeauthoryear{{Jin}, {Zhang}, {Zhang}, {Zhao}, {Wu}  \&
  {Fan}}{{Jin} et~al.}{2019}]{jin19}
{Jin} X.,  {Zhang} Y.,  {Zhang} J.,  {Zhao} Y.,  {Wu} X.-b.,   {Fan} D.,  2019,
  \mn@doi [\mnras] {10.1093/mnras/stz680}, \href
  {https://ui.adsabs.harvard.edu/abs/2019MNRAS.485.4539J} {485, 4539}

\bibitem[\protect\citeauthoryear{{Kallman}}{{Kallman}}{2001}]{kallman01}
{Kallman} T.~R.,  2001, in {Ferland} G.,  {Savin} D.~W.,  eds,  Astronomical
  Society of the Pacific Conference Series Vol. 247, Spectroscopic Challenges
  of Photoionized Plasmas. p.~175

\bibitem[\protect\citeauthoryear{{Kallman} \& {McCray}}{{Kallman} \&
  {McCray}}{1982}]{kallman_mccray82}
{Kallman} T.~R.,  {McCray} R.,  1982, \mn@doi [\apjs] {10.1086/190828}, \href
  {https://ui.adsabs.harvard.edu/abs/1982ApJS...50..263K} {50, 263}

\bibitem[\protect\citeauthoryear{{Kauffmann}, {White}  \&
  {Guiderdoni}}{{Kauffmann} et~al.}{1993}]{kauffmann93}
{Kauffmann} G.,  {White} S.~D.~M.,   {Guiderdoni} B.,  1993, \mn@doi [\mnras]
  {10.1093/mnras/264.1.201}, \href
  {https://ui.adsabs.harvard.edu/abs/1993MNRAS.264..201K} {264, 201}

\bibitem[\protect\citeauthoryear{{Kinkhabwala}, {Behar}, {Sako}, {Gu}, {Kahn}
  \& {Paerels}}{{Kinkhabwala} et~al.}{2003}]{kinkhabwala03}
{Kinkhabwala} A.,  {Behar} E.,  {Sako} M.,  {Gu} M.~F.,  {Kahn} S.~M.,
  {Paerels} F.~B.~S.,  2003, \mn@doi [arXiv e-prints]
  {10.48550/arXiv.astro-ph/0304332}, \href
  {https://ui.adsabs.harvard.edu/abs/2003astro.ph..4332K} {pp
  astro--ph/0304332}

\bibitem[\protect\citeauthoryear{{Kravtsov}}{{Kravtsov}}{2003}]{kravtsov03}
{Kravtsov} A.~V.,  2003, \mn@doi [\apjl] {10.1086/376674}, \href
  {https://ui.adsabs.harvard.edu/abs/2003ApJ...590L...1K} {590, L1}

\bibitem[\protect\citeauthoryear{{Kuns{\'a}gi-M{\'a}t{\'e}}, {Beck}, {Szapudi}
  \& {Csabai}}{{Kuns{\'a}gi-M{\'a}t{\'e}} et~al.}{2022}]{kunsagi-mate22}
{Kuns{\'a}gi-M{\'a}t{\'e}} S.,  {Beck} R.,  {Szapudi} I.,   {Csabai} I.,  2022,
  \mn@doi [\mnras] {10.1093/mnras/stac2411}, \href
  {https://ui.adsabs.harvard.edu/abs/2022MNRAS.516.2662K} {516, 2662}

\bibitem[\protect\citeauthoryear{{Leitherer} et~al.,}{{Leitherer}
  et~al.}{1999}]{starburst99}
{Leitherer} C.,  et~al., 1999, \mn@doi [\apjs] {10.1086/313233}, \href
  {https://ui.adsabs.harvard.edu/abs/1999ApJS..123....3L} {123, 3}

\bibitem[\protect\citeauthoryear{{Li} et~al.,}{{Li} et~al.}{2021}]{li21}
{Li} C.,  et~al., 2021, \mn@doi [\mnras] {10.1093/mnras/stab1650}, \href
  {https://ui.adsabs.harvard.edu/abs/2021MNRAS.506.1651L} {506, 1651}

\bibitem[\protect\citeauthoryear{{Lucey} et~al.,}{{Lucey}
  et~al.}{2023}]{lucey22}
{Lucey} M.,  et~al., 2023, \mn@doi [\mnras] {10.1093/mnras/stad1675}, \href
  {https://ui.adsabs.harvard.edu/abs/2023MNRAS.523.4049L} {523, 4049}

\bibitem[\protect\citeauthoryear{{Lundberg} \& {Lee}}{{Lundberg} \&
  {Lee}}{2017}]{lundberg_lee17}
{Lundberg} S.,  {Lee} S.-I.,  2017, \mn@doi [arXiv e-prints]
  {10.48550/arXiv.1705.07874}, \href
  {https://ui.adsabs.harvard.edu/abs/2017arXiv170507874L} {p. arXiv:1705.07874}

\bibitem[\protect\citeauthoryear{{Lundberg}, {Erion}  \& {Lee}}{{Lundberg}
  et~al.}{2018}]{lundberg18}
{Lundberg} S.~M.,  {Erion} G.~G.,   {Lee} S.-I.,  2018, \mn@doi [arXiv
  e-prints] {10.48550/arXiv.1802.03888}, \href
  {https://ui.adsabs.harvard.edu/abs/2018arXiv180203888L} {p. arXiv:1802.03888}

\bibitem[\protect\citeauthoryear{Lundberg et~al.,}{Lundberg
  et~al.}{2020}]{lundberg20}
Lundberg S.~M.,  et~al., 2020, Nature Machine Intelligence, 2, 2522

\bibitem[\protect\citeauthoryear{{Luo}, {Wang}, {Zhu-Ge}, {Li}, {Zou}  \&
  {Zhang}}{{Luo} et~al.}{2022}]{luo22}
{Luo} J.-W.,  {Wang} F.-F.,  {Zhu-Ge} J.-M.,  {Li} Y.,  {Zou} Y.-C.,   {Zhang}
  B.,  2022, \mn@doi [arXiv e-prints] {10.48550/arXiv.2211.16451}, \href
  {https://ui.adsabs.harvard.edu/abs/2022arXiv221116451L} {p. arXiv:2211.16451}

\bibitem[\protect\citeauthoryear{{Lykins}, {Ferland}, {Porter}, {van Hoof},
  {Williams}  \& {Gnat}}{{Lykins} et~al.}{2013}]{lykins13}
{Lykins} M.~L.,  {Ferland} G.~J.,  {Porter} R.~L.,  {van Hoof} P. A.~M.,
  {Williams} R.~J.~R.,   {Gnat} O.,  2013, \mn@doi [\mnras]
  {10.1093/mnras/sts570}, \href
  {https://ui.adsabs.harvard.edu/abs/2013MNRAS.429.3133L} {429, 3133}

\bibitem[\protect\citeauthoryear{{Machado Poletti Valle}, {Avestruz}, {Barnes},
  {Farahi}, {Lau}  \& {Nagai}}{{Machado Poletti Valle}
  et~al.}{2021}]{machado21}
{Machado Poletti Valle} L.~F.,  {Avestruz} C.,  {Barnes} D.~J.,  {Farahi} A.,
  {Lau} E.~T.,   {Nagai} D.,  2021, \mn@doi [\mnras] {10.1093/mnras/stab2252},
  \href {https://ui.adsabs.harvard.edu/abs/2021MNRAS.507.1468M} {507, 1468}

\bibitem[\protect\citeauthoryear{{Mart{\'\i}nez-Serrano}, {Serna},
  {Dom{\'\i}nguez-Tenreiro}  \& {Moll{\'a}}}{{Mart{\'\i}nez-Serrano}
  et~al.}{2008}]{martinez_serrano08}
{Mart{\'\i}nez-Serrano} F.~J.,  {Serna} A.,  {Dom{\'\i}nguez-Tenreiro} R.,
  {Moll{\'a}} M.,  2008, \mn@doi [\mnras] {10.1111/j.1365-2966.2008.13383.x},
  \href {https://ui.adsabs.harvard.edu/abs/2008MNRAS.388...39M} {388, 39}

\bibitem[\protect\citeauthoryear{{McCarthy}, {Schaye}, {Bird}  \& {Le
  Brun}}{{McCarthy} et~al.}{2017}]{mccarthy17}
{McCarthy} I.~G.,  {Schaye} J.,  {Bird} S.,   {Le Brun} A. M.~C.,  2017,
  \mn@doi [\mnras] {10.1093/mnras/stw2792}, \href
  {https://ui.adsabs.harvard.edu/abs/2017MNRAS.465.2936M} {465, 2936}

\bibitem[\protect\citeauthoryear{{Mirabal}, {Charles}, {Ferrara}, {Gonthier},
  {Harding}, {S{\'a}nchez-Conde}  \& {Thompson}}{{Mirabal}
  et~al.}{2016}]{mirabal16}
{Mirabal} N.,  {Charles} E.,  {Ferrara} E.~C.,  {Gonthier} P.~L.,  {Harding}
  A.~K.,  {S{\'a}nchez-Conde} M.~A.,   {Thompson} D.~J.,  2016, \mn@doi [\apj]
  {10.3847/0004-637X/825/1/69}, \href
  {https://ui.adsabs.harvard.edu/abs/2016ApJ...825...69M} {825, 69}

\bibitem[\protect\citeauthoryear{{Morisset}, {Stasi{\'n}ska}  \&
  {Pe{\~n}a}}{{Morisset} et~al.}{2005}]{morisset05}
{Morisset} C.,  {Stasi{\'n}ska} G.,   {Pe{\~n}a} M.,  2005, \mn@doi [\mnras]
  {10.1111/j.1365-2966.2005.09049.x}, \href
  {https://ui.adsabs.harvard.edu/abs/2005MNRAS.360..499M} {360, 499}

\bibitem[\protect\citeauthoryear{{Nakoneczny} et~al.,}{{Nakoneczny}
  et~al.}{2021}]{nakoneczny21}
{Nakoneczny} S.~J.,  et~al., 2021, \mn@doi [\aap]
  {10.1051/0004-6361/202039684}, \href
  {https://ui.adsabs.harvard.edu/abs/2021A&A...649A..81N} {649, A81}

\bibitem[\protect\citeauthoryear{{Okamoto}, {Gao}  \& {Theuns}}{{Okamoto}
  et~al.}{2008}]{okamoto08}
{Okamoto} T.,  {Gao} L.,   {Theuns} T.,  2008, \mn@doi [\mnras]
  {10.1111/j.1365-2966.2008.13830.x}, \href
  {https://ui.adsabs.harvard.edu/abs/2008MNRAS.390..920O} {390, 920}

\bibitem[\protect\citeauthoryear{Pedregosa et~al.,}{Pedregosa
  et~al.}{2011}]{scikit-learn}
Pedregosa F.,  et~al., 2011, Journal of Machine Learning Research, 12, 2825

\bibitem[\protect\citeauthoryear{Ploeckinger \& Schaye}{Ploeckinger \&
  Schaye}{2020}]{ploeckinger_schaye20}
Ploeckinger S.,  Schaye J.,  2020, \mn@doi [Monthly Notices of the Royal
  Astronomical Society] {10.1093/mnras/staa2172}, 497, 4857

\bibitem[\protect\citeauthoryear{{Rees} \& {Ostriker}}{{Rees} \&
  {Ostriker}}{1977}]{rees_ostriker77}
{Rees} M.~J.,  {Ostriker} J.~P.,  1977, \mn@doi [\mnras]
  {10.1093/mnras/179.4.541}, \href
  {https://ui.adsabs.harvard.edu/abs/1977MNRAS.179..541R} {179, 541}

\bibitem[\protect\citeauthoryear{{Richings}, {Schaye}  \&
  {Oppenheimer}}{{Richings} et~al.}{2014}]{richings14}
{Richings} A.~J.,  {Schaye} J.,   {Oppenheimer} B.~D.,  2014, \mn@doi [\mnras]
  {10.1093/mnras/stu525}, \href
  {https://ui.adsabs.harvard.edu/abs/2014MNRAS.440.3349R} {440, 3349}

\bibitem[\protect\citeauthoryear{{Robertson} \& {Kravtsov}}{{Robertson} \&
  {Kravtsov}}{2008}]{robertson_kravtsov08}
{Robertson} B.~E.,  {Kravtsov} A.~V.,  2008, \mn@doi [\apj] {10.1086/587796},
  \href {https://ui.adsabs.harvard.edu/abs/2008ApJ...680.1083R} {680, 1083}

\bibitem[\protect\citeauthoryear{{Robinson}, {Avestruz}  \&
  {Gnedin}}{{Robinson} et~al.}{2022}]{robinson22}
{Robinson} D.,  {Avestruz} C.,   {Gnedin} N.~Y.,  2022, \mn@doi [\apj]
  {10.3847/1538-4357/ac85e1}, \href
  {https://ui.adsabs.harvard.edu/abs/2022ApJ...936...50R} {936, 50}

\bibitem[\protect\citeauthoryear{{Romero}, {Ascasibar}, {Palou{\v{s}}},
  {W{\"u}nsch}  \& {Moll{\'a}}}{{Romero} et~al.}{2021}]{romero21}
{Romero} M.,  {Ascasibar} Y.,  {Palou{\v{s}}} J.,  {W{\"u}nsch} R.,
  {Moll{\'a}} M.,  2021, \mn@doi [\mnras] {10.1093/mnras/stab1660}, \href
  {https://ui.adsabs.harvard.edu/abs/2021MNRAS.505.5301R} {505, 5301}

\bibitem[\protect\citeauthoryear{{Salz}, {Banerjee}, {Mignone}, {Schneider},
  {Czesla}  \& {Schmitt}}{{Salz} et~al.}{2015}]{salz15}
{Salz} M.,  {Banerjee} R.,  {Mignone} A.,  {Schneider} P.~C.,  {Czesla} S.,
  {Schmitt} J.~H.~M.~M.,  2015, \mn@doi [\aap] {10.1051/0004-6361/201424330},
  \href {https://ui.adsabs.harvard.edu/abs/2015A&A...576A..21S} {576, A21}

\bibitem[\protect\citeauthoryear{Schaye et~al.,}{Schaye
  et~al.}{2014}]{schaye14}
Schaye J.,  et~al., 2014, \mn@doi [Monthly Notices of the Royal Astronomical
  Society] {10.1093/mnras/stu2058}, 446, 521

\bibitem[\protect\citeauthoryear{{Schaye} et~al.,}{{Schaye}
  et~al.}{2023}]{schaye23}
{Schaye} J.,  et~al., 2023, \mn@doi [\mnras] {10.1093/mnras/stad2419}, \href
  {https://ui.adsabs.harvard.edu/abs/2023MNRAS.tmp.2384S} {}

\bibitem[\protect\citeauthoryear{Shwartz-Ziv \& Armon}{Shwartz-Ziv \&
  Armon}{2022}]{shwartz-ziv_armon21}
Shwartz-Ziv R.,  Armon A.,  2022, Information Fusion, 81, 84

\bibitem[\protect\citeauthoryear{{Silk}}{{Silk}}{1977}]{silk77}
{Silk} J.,  1977, \mn@doi [\apj] {10.1086/154972}, \href
  {https://ui.adsabs.harvard.edu/abs/1977ApJ...211..638S} {211, 638}

\bibitem[\protect\citeauthoryear{{Smith}, {Sigurdsson}  \& {Abel}}{{Smith}
  et~al.}{2008}]{smith08}
{Smith} B.,  {Sigurdsson} S.,   {Abel} T.,  2008, \mn@doi [\mnras]
  {10.1111/j.1365-2966.2008.12922.x}, \href
  {https://ui.adsabs.harvard.edu/abs/2008MNRAS.385.1443S} {385, 1443}

\bibitem[\protect\citeauthoryear{{Smith} et~al.,}{{Smith}
  et~al.}{2017}]{smith17}
{Smith} B.~D.,  et~al., 2017, \mn@doi [\mnras] {10.1093/mnras/stw3291}, \href
  {https://ui.adsabs.harvard.edu/abs/2017MNRAS.466.2217S} {466, 2217}

\bibitem[\protect\citeauthoryear{Spitzer}{Spitzer}{1962}]{spitzer62}
Spitzer L.,  1962, Physics of fully ionized gases, 2d rev. ed edn.
Interscience Publishers New York, New York

\bibitem[\protect\citeauthoryear{{Sutherland} \& {Dopita}}{{Sutherland} \&
  {Dopita}}{1993}]{sutherland_dopita93}
{Sutherland} R.~S.,  {Dopita} M.~A.,  1993, \mn@doi [\apjs] {10.1086/191823},
  \href {https://ui.adsabs.harvard.edu/abs/1993ApJS...88..253S} {88, 253}

\bibitem[\protect\citeauthoryear{{Tamayo} et~al.,}{{Tamayo}
  et~al.}{2016}]{tamayo16}
{Tamayo} D.,  et~al., 2016, \mn@doi [\apjl] {10.3847/2041-8205/832/2/L22},
  \href {https://ui.adsabs.harvard.edu/abs/2016ApJ...832L..22T} {832, L22}

\bibitem[\protect\citeauthoryear{{Thomas} et~al.,}{{Thomas}
  et~al.}{2009}]{thomas09}
{Thomas} R.~M.,  et~al., 2009, \mn@doi [\mnras]
  {10.1111/j.1365-2966.2008.14206.x}, \href
  {https://ui.adsabs.harvard.edu/abs/2009MNRAS.393...32T} {393, 32}

\bibitem[\protect\citeauthoryear{{Vogelsberger} et~al.,}{{Vogelsberger}
  et~al.}{2014}]{vogelsberger14}
{Vogelsberger} M.,  et~al., 2014, \mn@doi [\nat] {10.1038/nature13316}, \href
  {https://ui.adsabs.harvard.edu/abs/2014Natur.509..177V} {509, 177}

\bibitem[\protect\citeauthoryear{{Wang}, {Ferland}, {Lykins}, {Porter}, {van
  Hoof}  \& {Williams}}{{Wang} et~al.}{2014}]{wang14}
{Wang} Y.,  {Ferland} G.~J.,  {Lykins} M.~L.,  {Porter} R.~L.,  {van Hoof}
  P.~A.~M.,   {Williams} R.~J.~R.,  2014, \mn@doi [\mnras]
  {10.1093/mnras/stu514}, \href
  {https://ui.adsabs.harvard.edu/abs/2014MNRAS.440.3100W} {440, 3100}

\bibitem[\protect\citeauthoryear{{Wang}, {Pan}, {Zheng}, {Qian}  \&
  {Li}}{{Wang} et~al.}{2019}]{wang19}
{Wang} Y.,  {Pan} Z.,  {Zheng} J.,  {Qian} L.,   {Li} M.,  2019, \mn@doi
  [\apss] {10.1007/s10509-019-3602-4}, \href
  {https://ui.adsabs.harvard.edu/abs/2019Ap&SS.364..139W} {364, 139}

\bibitem[\protect\citeauthoryear{{White} \& {Frenk}}{{White} \&
  {Frenk}}{1991}]{white_frenk91}
{White} S. D.~M.,  {Frenk} C.~S.,  1991, \mn@doi [\apj] {10.1086/170483}, \href
  {https://ui.adsabs.harvard.edu/abs/1991ApJ...379...52W} {379, 52}

\bibitem[\protect\citeauthoryear{{Wiersma}, {Schaye}  \& {Smith}}{{Wiersma}
  et~al.}{2009}]{wiersma09}
{Wiersma} R. P.~C.,  {Schaye} J.,   {Smith} B.~D.,  2009, \mn@doi [\mnras]
  {10.1111/j.1365-2966.2008.14191.x}, \href
  {https://ui.adsabs.harvard.edu/abs/2009MNRAS.393...99W} {393, 99}

\bibitem[\protect\citeauthoryear{{Wiersma}, {Schaye}, {Dalla Vecchia}, {Booth},
  {Theuns}  \& {Aguirre}}{{Wiersma} et~al.}{2010}]{wiersma10}
{Wiersma} R. P.~C.,  {Schaye} J.,  {Dalla Vecchia} C.,  {Booth} C.~M.,
  {Theuns} T.,   {Aguirre} A.,  2010, \mn@doi [\mnras]
  {10.1111/j.1365-2966.2010.17299.x}, \href
  {https://ui.adsabs.harvard.edu/abs/2010MNRAS.409..132W} {409, 132}

\bibitem[\protect\citeauthoryear{{Wood}, {Mathis}  \& {Ercolano}}{{Wood}
  et~al.}{2004}]{wood04}
{Wood} K.,  {Mathis} J.~S.,   {Ercolano} B.,  2004, \mn@doi [\mnras]
  {10.1111/j.1365-2966.2004.07458.x}, \href
  {https://ui.adsabs.harvard.edu/abs/2004MNRAS.348.1337W} {348, 1337}

\bibitem[\protect\citeauthoryear{Zebari, Abdulazeez, Zeebaree, Zebari  \&
  Saeed}{Zebari et~al.}{2020}]{zebari20}
Zebari R.,  Abdulazeez A.,  Zeebaree D.,  Zebari D.,   Saeed J.,  2020, \mn@doi
  [Journal of Applied Science and Technology Trends] {10.38094/jastt1224}, 1,
  56

\makeatother
\end{thebibliography}



\appendix

\section{Hyperparameter Tuning}
\label{app:hyperparams}

There are many hyperparameters that can be tuned to adjust how precisely an XGBoost model can be fit to its training data\footnote{Detailed descriptions of these hyperparameters can be found in the XGBoost documentation: \url{https://xgboost.readthedocs.io/en/stable/index.html}}.  For this paper, we consider non-default values for only 7 of the available hyperparameters:
\begin{itemize}
    \item \texttt{max\_depth}: the maximum depth of each tree.  Default: 6, must be a positive integer. 
    \item \texttt{min\_child\_weight}: the minimum weight required to split a node, where smaller means a more complex model. Default: 1, must be non-negative. 
    \item \texttt{subsample}: the fraction of training data rows used in each tree.  Default: 1, must be between 0 and 1 (inclusive). 
    \item \texttt{colsample\_bytree}: the fraction of features used in each tree. Default: 1, must be between 0 and 1 (inclusive). 
    \item \texttt{gamma}: the cost for adding each new node in the objective function. Default: 0, must be non-negative. 
    \item \texttt{eta}: the learning rate.  Default: 0.3, must be positive.
    \item \texttt{n\_estimators}: the number of trees in the ensemble (results of each tree are added together). Since the trees are trained sequentially, this is equivalent to the number of training rounds.  Default: 100, must be a positive integer.
\end{itemize}

To find the `best' hyperparameters using the validation set (described in Section~\ref{method:model_training}), we use 5-fold cross validation, as implemented in \texttt{scikit-learn} and the Bayesian optimization scheme \texttt{BayesSearchCV} from \texttt{scikit-optimize} \citep{scikit-optimize}.  We set up the search space as follows.
\begin{itemize}
    \item \texttt{min\_child\_weight}: Real, log-uniform prior between 0.1 and 2. 
    \item \texttt{subsample}: Real, uniform prior between 0.6 and 1.
    \item \texttt{colsample\_bytree}: Real, uniform prior between 0.6 and 1.
    \item \texttt{gamma}: Real, uniform prior between 0 and 1.
    \item \texttt{eta}: Real, log-uniform prior between 0.03 and 0.3.
\end{itemize}
We use the negative mean squared error as the scoring method for the Bayesian hyperparameter search.

In principle, the error on the training set (defined in Section~\ref{method:model_training}) can be made arbitrary close to 0 with an overfitted model.  However, we do not necessarily need to find the lowest possible error, but rather are looking for a sufficiently well-performing set of hyperparameters. For these reasons, rather than incorporating them into the Bayesian search, we perform a manual grid search in \texttt{max$\_$depth} and \texttt{n$\_$estimators} (the two hyperparameters which we found to have the strongest effect on model performance), where we explore the Cartesian product of:
\begin{align*}
    \texttt{max\_depth} &= \{8, 12, 16, 20\}, \\
    \texttt{n\_estimators} &= \{1, 10, 100, 300, 1000\}.
\end{align*}
At each of these 40 combinations of $\{$\texttt{max$\_$depth}, \texttt{n$\_$estimators}$\}$, we perform the Bayesian optimization of the other 5 hyperparameters as described above. We compare both the mean squared error and the cumulative error distribution evaluated at $\Delta \log \mathcal{F}=0.3$, $P(\Delta \log \mathcal{F} > 0.3)$ (the frequency of errors $\Delta \log \mathcal{F} > 0.3$) on the test set, to constrain the frequency of large errors.  The latter quantity is more meaningful for discriminating between similar well-performing models, because the mean squared error is dominated by the more frequent smaller errors and is not very sensitive to changes in the magnitude of the largest, most infrequent errors. As an additional comparison, we evaluate the time needed to train each of the 40 models.  In general, pairs $\{$\texttt{max$\_$depth}, \texttt{n$\_$estimators}$\}$ which produce lower mean squared error and/or frequency of large errors on the test set have longer training times.

\begin{figure*}
    \centering
    \includegraphics[width = 2\columnwidth]{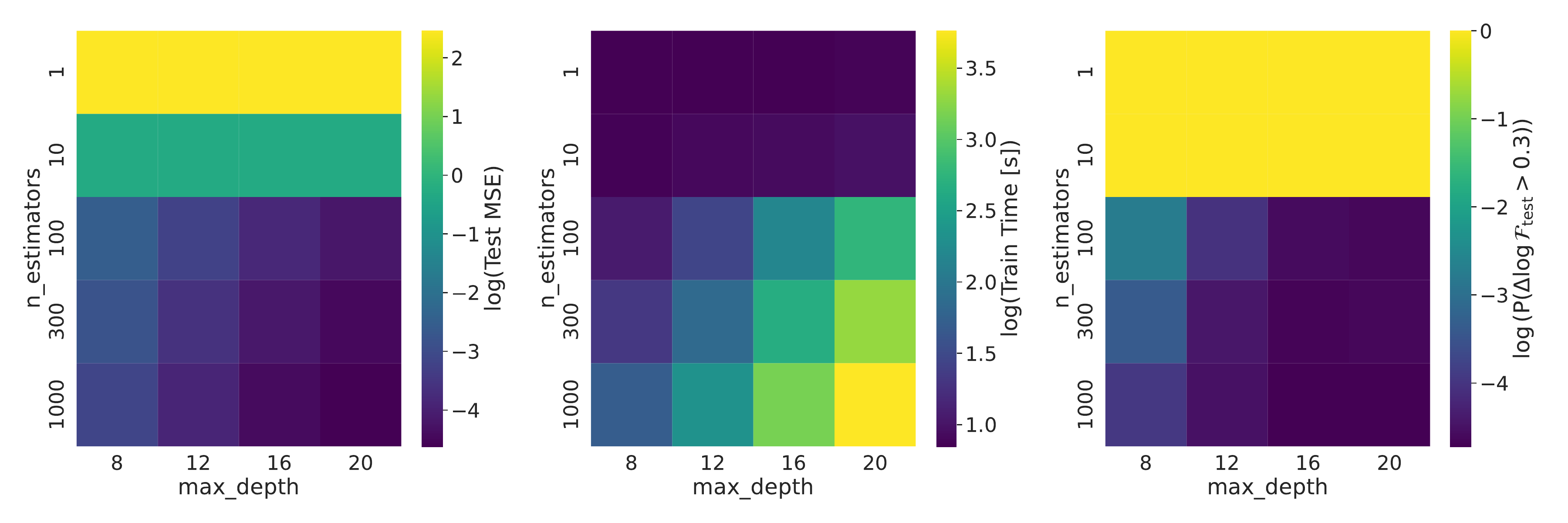}
    \caption{Results from a grid search in \texttt{max\_depth} and \texttt{n\_estimators} for cooling function model trained with the 21 scaled rates from the PCA of section~\ref{method:pca_rates} at $Z/Z_\odot = 1$.  The three panels show heatmaps of the mean squared error in $\log\mathcal{F}$ on the test set (a randomly select 20\% of the training data table described in Section~\ref{method:data} (left), the training time in seconds (middle), and the frequency of errors $\Delta \log \mathcal{F}$ larger than $0.3$ on the test set (right).  The heatmaps are all on a logarithmic scale.}
    \label{fig:grid_search_ex}
\end{figure*}

An example of such a grid search, for a cooling function model trained using the 21 scaled rates selected by the PCA described in section~\ref{method:pca_rates} at $Z/Z_\odot = 1$ is shown in Fig.~\ref{fig:grid_search_ex}.  From this grid search, we selected $\texttt{max\_depth}=20, \texttt{n\_estimators}=100$ as our choice, because it has the shortest training time (middle panel) among choices of \texttt{max\_depth} and \texttt{n\_estimators} with similar mean squared error and frequency of large errors (left- and right- most panels in Fig.~\ref{fig:grid_search_ex} to the best cases in the grid search.

We find that the results of such grid searches, and in particular the choice of $\texttt{max\_depth}=20, \texttt{n\_estimators}=100$, are consistent across models trained using the 21 scaled rates identified in the PCA analysis described in section~\ref{method:pca_rates} for both cooling and heating functions at $Z/Z_\odot = 0$ and $1$, and using both the 21 scaled rates from Section~\ref{method:pca_rates} and only the top 3 scaled rates from the feature importance analysis discussed in Section~\ref{method:shap}.
 
After confirming that the choice of $\texttt{max\_depth}=20$,  $\texttt{n\_estimators}=100$ is robust across metallicity, cooling vs. heating functions, and for models with very different numbers of scaled rate features, we adopted \texttt{max\_depth}$=20,$ \texttt{n\_estimators}$=100$ for all other models, without performing further grid searches in \texttt{max\_depth} and \texttt{n\_estimators} (but still performing a Bayesian optimization of the other 5 hyperparameters we consider). 

\section{Comparison on training data for all metallicities}
\label{app:all_Z_training_comp}

\begin{figure*}
    \centering
    \includegraphics[width = \textwidth]{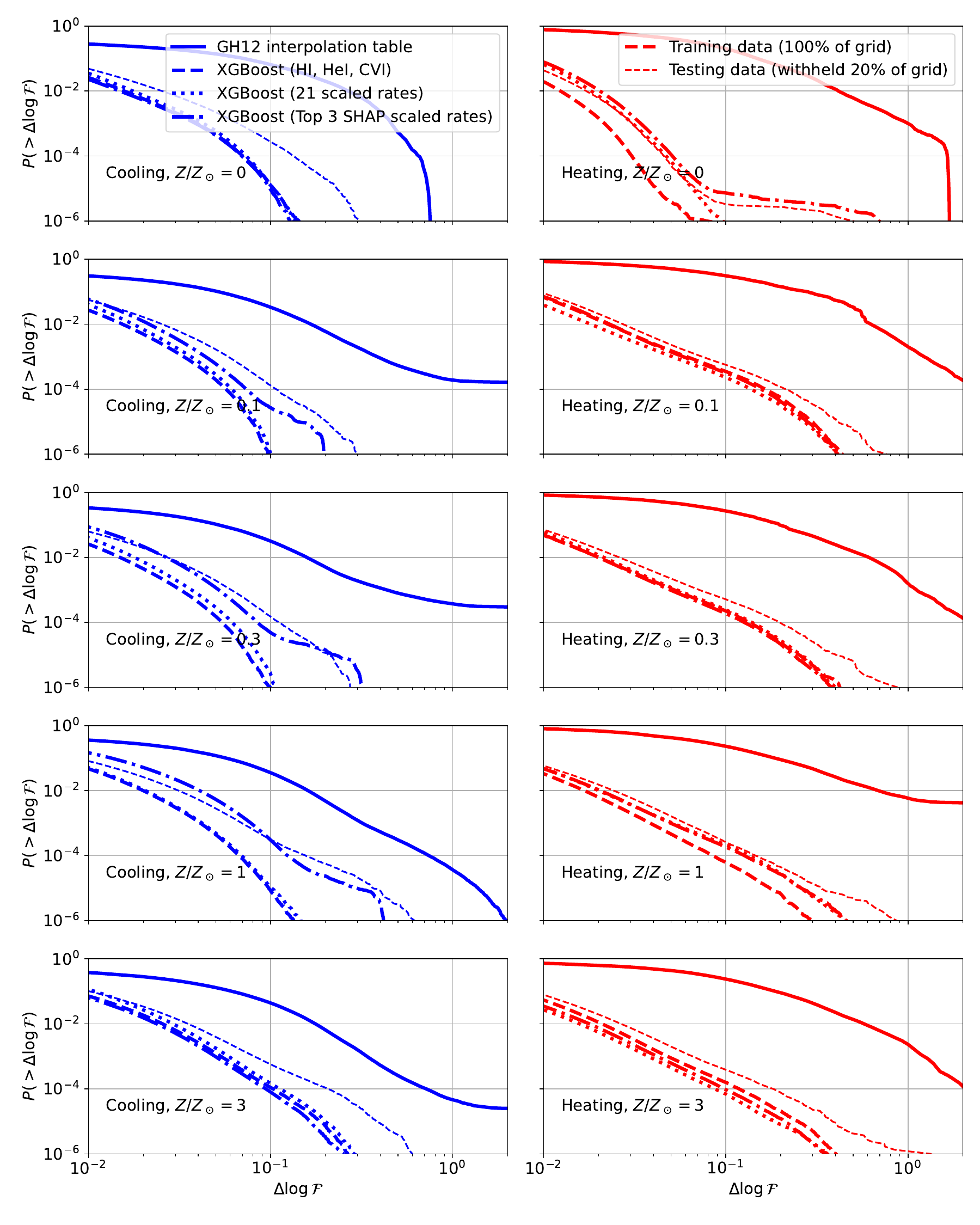}
    \caption{Same as Fig.~\ref{fig:fixed_Z_cdfs}, but now for all 5 metallicity values in the training data table (described in Table~\ref{tab:training_data}).}
    \label{fig:all_z_fixed_z_cdfs}
\end{figure*}

In Fig.~\ref{fig:all_z_fixed_z_cdfs}, we extend the comparison in Fig.~\ref{fig:fixed_Z_cdfs} to all metallicities in the training data table (described in Table~\ref{tab:training_data}): $Z/Z_\odot = \{0, 0.1, 0.3, 1,3 \}$.


\bsp	
\label{lastpage}
\end{document}